\documentclass{jfm}
\usepackage[pdftex]{graphicx}
\usepackage{psfrag, mathrsfs, subfig, amsmath, amssymb, natbib, multirow, bbding}
\usepackage{epstopdf}

\DeclareFontFamily{OT1}{pzc}{}
\DeclareFontShape{OT1}{pzc}{m}{it}{<-> s * [1.10] pzcmi7t}{}
\DeclareMathAlphabet{\mathpzc}{OT1}{pzc}{m}{it}

\newcommand\pd[2]{\frac{\partial#1}{\partial#2}}
\newcommand\vc[1]{\boldsymbol{#1}}
\newcommand\dv[2]{\frac{\ud #1}{\ud #2}}
\newcommand\Dv[2]{\frac{\textrm{D} #1}{\textrm{D} #2}}
\newcommand\ud{\textrm{d}}

\newcommand\tn[1]{\underline{\underline{#1}}}
\newcommand\Rey{\textrm{Re}}
\newcommand\We{\textrm{We}}
\newcommand\Pe{\textrm{Pe}}

\makeatletter
\newcommand\Rmnum[1]{\expandafter\@slowromancap\romannumeral #1@}
\makeatother

\title[Centrifugally forced Rayleigh-Taylor Instability]
  {Centrifugally forced Rayleigh-Taylor Instability}

\author[M.~M.~Scase \& R.~J.~A.~Hill]
{M.\ns M.\ns S\ls C\ls A\ls S\ls E$^1$ \& R.\ns J.\ns A.\ns H\ls I\ls L\ls L$^2$}

\affiliation{$^1$School of Mathematical Sciences, University of Nottingham, Nottingham NG7 2RD, UK\\
                $^2$School of Physics and Astronomy, University of Nottingham, Nottingham NG7 2RD, UK}

\date{\today} 

\begin{document}

\maketitle

\begin{abstract}
The effect of rotation on the classical gravity-driven Rayleigh-Taylor instability has been shown to influence the scale of the perturbations that develop at the unstable interface and consequently alter the speed of propagation of the front.  The present authors argued that this is as a result of a competition between the destabilizing effect of gravity and the stabilizing effect of the rotation.  The case considered was for reasonably low rotation rates applied to statically unstable layers of fluid in a cylindrical geometry where the interface adopts a parabolic profile.  In the present paper we consider the extreme limit of high rotation rates in which rotational forces dominate and gravitational forces may be ignored.  The two liquid layers initially form concentric cylinders, centred on the axis of rotation.  The configuration may be thought of as a fluid-fluid centrifuge. There are two types of perturbation to the interface that may be considered, an azimuthal perturbation around the circumference of the interface and a varicose perturbation in the axial direction along the length of the interface.  It is the first of these types of perturbation that we consider here, and so the flow may be considered essentially two-dimensional, taking place in a circular domain.

We carry out a linear stability analysis on a perturbation to the hydrostatic background state and derive a fourth order Orr-Sommerfeld-like equation that governs the system.  We consider the dynamics of systems of stable and unstable configurations, inviscid and viscous fluids, immiscible fluid layers with surface tension, and miscible fluid layers that may have some initial diffusion of density.  Theoretical predictions are compared with numerical experiments and the agreement is shown to be good.  We do not restrict our analysis to equal volume fluid layers and so our results also have applications in coating and lubrication problems in rapidly rotating systems and machinery.
\end{abstract}


\section{\label{sec:int}Introduction}


The effects of rotation on the Rayleigh-Taylor instability \citep{rayleigh83, taylor50} have been considered by a number of different authors, from the theoretical work of \citet{hide56} and \citet{chandra}, for example, to later numerical studies by \citet{carnevaleEtAl} and \citet{boffettaetal} and more recent experimental \citep[see e.g.,][]{scirep, jove} and analytical advances \citep[see e.g.,][]{tao, prf}.  In all cases the effect of rotation on the fundamental gravitational instability caused by a dense fluid lying above a less dense fluid has been considered.  The effect of the rotation on the system is to form so-called Taylor columns \citep[][]{taylor23} in the density-stratified system that inhibit the formation of large-scale eddies at the interface.  This in turn inhibits the development and propagation speed of the interface and so the rate and scale of the instability can be controlled to some extent by the rotation.

\begin{figure}
\begin{center}
\includegraphics{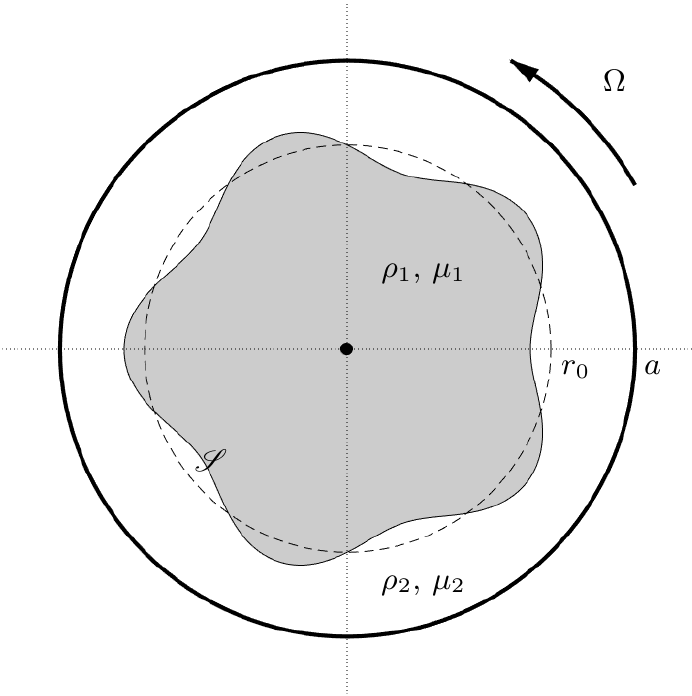}
\end{center}
\caption{\label{fig:schematic}Schematic of the two-dimensional circular flow domain.  The system is rotating with angular frequency $\Omega$ and has radius $a$.  The fluid density and dynamic viscosity in the inner and outer layers are $\rho_1$, $\mu_1$ and $\rho_2$, $\mu_2$ respectively.  The initial radius of the inner layer is $r_0\in(0,a)$.  The interface is denoted by $\mathscr{S}$.}
\end{figure}

Here we consider the limiting case of a rotating system in the absence of a gravitational field.  This may be considered the high-rotation rate limit of the rotating Rayleigh-Taylor problem.   In the high-rotation limit the parabolic interface that occurs at the interface between two fluids at low rotation rates becomes a cylindrical interface between two concentric cylinders of fluid aligned on the axis of rotation.  If the outer layer of fluid is denser than the inner layer then the system is stable and supports interfacial waves.  Conversely, if the outer layer is less dense than the inner layer then the system is unstable and a perturbation to the interface may grow in time as the system seeks a more stable configuration.  The system supports perturbations in both the azimuthal and axial directions and we focus our attention upon the azimuthal perturbations here.  This restricts the flow to a two-dimensional plane polar coordinate system that we consider in a circular domain.  As the considered flow is strictly two-dimensional Taylor-Couette flow is prohibited \citep[see e.g.,][]{pengZhu}.  The set up is shown in figure \ref{fig:schematic} for a perturbation with azimuthal wavenumber 5.

We consider a number of possible scenarios depending on whether the fluids are viscous or inviscid, immiscible with surface tension, or miscible with and without an initially diffuse interface.  We do not restrict the analysis to equal fluid volumes in each layer.  The reason for the choices of different scenarios is to develop results that mirror the well-established results in the literature on classical Rayleigh-Taylor instability \citep[see e.g.,][]{chandra}.  We find that surface tension can enhance the frequency of stable modes of oscillation and, as with classical Rayleigh-Taylor instability, can act to stabilize otherwise unstable modes of instability.  In the case of a diffuse interface there exists a band of fluid of radially varying density that can support a number of inertial waves.  The wider the diffuse layer, or, equivalently, the higher the azimuthal wavenumber, the greater the number of inertial waves that can be supported.

In the most simple case of two inviscid fluid layers separated by a sharp interface with no diffusion or surface tension we are able to show that the growth of perturbations in an unstable configuration is due to the centrifugal term in the governing equation of motion.  The Coriolis term can modify the growth, but is unable to alter the stability of the system.  This can be observed from the fact that the dispersion relation is a quadratic expression for the growth rate and the Coriolis term can only make a positive contribution to the discriminant, whereas the centrifugal term makes a contribution whose sign depends accordingly on the stability or instability of the system.

The flow we are investigating is similar to that studied by \citet{tao} who considered the effects of rotation on concentric cylinders of density-stratified inviscid fluid with the additional feature that the interface is accelerated toward the lighter fluid.  Our analysis differs in that we begin from a hydrostatic configuration, we do not ignore the effects of curvature on the interface (i.e., we do not assume that the azimuthal wavelength is negligible compared to the radius of the inner fluid), nor do we assume that the flow is taking place in an unbounded domain -- we specifically consider the importance of the ratio of volumes of fluid.  The unbounded domain is a limit of our problem and in the inviscid, non-diffuse, zero surface tension case we demonstrate that, in the appropriate limits, we recover the dispersion relation of \citet{tao}.

In the case of viscous fluid layers, we observe that the key consideration is the viscosity of the most viscous layer and that this layer dominates the behaviour of the system.  We also look at the high viscosity, zero-inertia limit of our system by making a Stokes flow approximation.  We compare our results to those of \citet{schwartz} and \citet{alvarez-lacalle} who considered the effects of rotation on concentric cylinders of fluid of varying viscosity in Hele-Shaw cells and porous media.  We show that the underlying physics between the two systems is similar but predictions differ by a shape factor due to the contrasting ways that shearing affects the two models.

The structure of the paper is as follows: in \S\,\ref{sec:model} we introduce the general governing model, allowing for radial variation of the fluid density and dynamic viscosity in each fluid layer.  We then consider the appropriate boundary conditions for miscible fluids with a sharp non-diffuse interface or immiscible fluids with surface tension, and we briefly describe the numerical techniques used for simulating the flows.  In \S\,\ref{sec:theComp} we consider our general set-up in a number of specific configurations and compare our predictions with numerical simulation.  In particular, we consider two uniform layers of inviscid fluid with a sharp interface (no diffusion) and no surface tension; this case is the rotational counter-part of the classical Rayleigh-Taylor instability work of \citet{rayleigh83} and \citet{taylor50}.  We then consider the effects of surface tension, the effects of diffusion of the interface prior to the onset of the instability, the effects of viscosity with and without surface tension acting and also, separately, with a diffuse interface.  Finally, we compare our results with those established for similar flows in Hele-Shaw cells and porous media.  In \S\,\ref{sec:conc} we discuss our results and draw our conclusions.


\section{\label{sec:model}Governing model}


\subsection{\label{sec:eqMotion}Equations of motion}


The governing equations for the fluid velocity, $\vc{u}_j$, pressure, $p_j$, density, $\rho_j$, and viscosity, $\mu_j$, in each fluid layer are the conservation of mass equation, an equation of motion and an incompressibility condition given respectively by
\begin{subequations}
\begin{equation}\label{eq:massCons1}
\pd{\rho_j}{t} + \nabla\cdot\left(\rho_j\vc{u}_j\right) = 0,
\end{equation}
\begin{equation} \label{eq:Mom1}
\rho_j\Dv{\vc{u}_j}{t} = -\nabla p_j - \rho_j\vc{\Omega}\times\left(\vc{\Omega}\times\vc{x}\right) - 2\rho_j\vc{\Omega}\times\vc{u}_j + \mu_j\nabla^2\vc{u}_j + 2\tn{e}_j\cdot\nabla\mu_j + \mu_j\nabla(\nabla\cdot\vc{u}_j),
\end{equation}
\begin{equation}\label{eq:inc1}
\Dv{\rho_j}{t}=0,
\end{equation}
\end{subequations}
for $j = 1, 2$ corresponding to the inner and outer layers respectively where $\tn{e}_j = \frac{1}{2}(\nabla\vc{u}_j + \nabla\vc{u}_j^\mathsf{T})$ is the rate of strain tensor in each layer. Equations \eqref{eq:massCons1} and \eqref{eq:inc1} combine to give the usual condition that $\nabla\cdot\vc{u}_j=0$ in each layer, removing the final term in \eqref{eq:Mom1}.
We nondimensionalise time by the angular velocity of the system, $\Omega$, and length by the radial extent of the domain, $a$.  The density and viscosity of the fluids in each layer are nondimensionalised by characteristic densities and viscosities of the whole system, respectively $\rho_0 = \frac{1}{2}(\rho_1+\rho_2)$ and $\mu_0 = \frac{1}{2}(\mu_1+\mu_2)$. The nondimensional system of governing equations is then
\begin{subequations} \label{eq:nond1}
\begin{equation}
\pd{\rho_j'}{t'} + \nabla'\cdot\left(\rho'_j\vc{u}'_j\right) = 0,
\end{equation}
\begin{equation}
\Dv{\vc{u}'_j}{t'} = -\frac{1}{\rho'_j}\nabla' p'_j +r'\hat{\vc{r}} - 2\hat{\vc{z}}\times\vc{u}'_j + \frac{1}{\Rey}\frac{\mu'_j}{\rho'_j}\nabla'^2\vc{u}'_j + \frac{2}{\Rey}\tn{e}_j'\cdot\nabla'\mu',
\end{equation}
\begin{equation}\label{eq:inc2}
\nabla'\cdot\vc{u}'_j=0,
\end{equation}
\end{subequations}
where the system Reynolds number is $\Rey = \rho_0 \Omega a^2/\mu_0$, the system pressure scale is $\rho_0 \Omega^2a^2$ and $\hat{\vc{r}}$ and $\hat{\vc{z}}$ are unit vectors in the radial and axial directions respectively.   As a result of the choice of velocity scale, the Reynolds number may be interpreted as a reciprocal Ekman number, where the Ekman number, describing the ratio of viscous forces to Coriolis forces, is given by $\textrm{Ek} = \mu_0/(\rho_0 \Omega a^2)$.  The prime symbols for nondimensional quantities are now dropped for clarity.

Hydrostatic solutions to the governing system \eqref{eq:nond1} are denoted by a superscript `$*$' and are given by
\begin{equation} \label{eq:SubsStart}
\vc{u}^*_j = \vc{0},
\quad
\rho^*_j = \rho^*_j(r),
\quad
p^*_j = p_0 + \int_0^r \rho^*_j(\xi)\xi\,\textrm{d}\xi,
\end{equation}
for a reference pressure $p_0$ at the origin and a radially varying initial density distribution $\rho^*_j = \rho^*_j(r)$.  In the hydrostatic configuration the form of $\mu_j$ is arbitrary.

We consider linear perturbations to the hydrostatic initial condition of the form
\begin{subequations} \label{eq:pertEqs}
\begin{gather}
\vc{u}_j = \vc{u}^*_j + \epsilon\vc{U}_j(r, \theta, t),
\quad
\rho_j = \rho^*_j(r) + \epsilon\sigma_j(r)\textrm{e}^{\textrm{i}(m\theta+\omega t)},
\\
p_j = p^*_j + \epsilon P_j(r)\textrm{e}^{\textrm{i}(m\theta+\omega t)},
\quad
\mu_j = \mu_j^*(r) + \epsilon\eta_j(r)\textrm{e}^{\textrm{i}(m\theta+\omega t)},
\end{gather}
\end{subequations}
where $\epsilon \ll 1$, $m\in\mathbb{N}$ and we allow for a fluid whose density and viscosity vary radially at leading order.  As the flow is incompressible, we may introduce a streamfunction $\psi$ such that
\begin{equation}
\vc{U}_j = \left(\frac{1}{r}\pd{\psi_j}{\theta}, -\pd{\psi_j}{r}, 0\right),
\end{equation}
and \eqref{eq:inc2} is automatically satisfied in each layer.  Writing the streamfunction as $\psi_j = \phi_j(r)\textrm{e}^{\textrm{i}(m\theta+\omega t)}$ we have
\begin{equation} \label{eq:ueq}
\vc{u}_j = \epsilon\left(\frac{\textrm{i}m\phi_j}{r}, -\dv{\phi_j}{r}, 0\right)\textrm{e}^{\textrm{i}(m\theta+\omega t)}.
\end{equation}
The solution growth rate is controlled by $\textrm{Im}(\omega)$, its precession is controlled by $\textrm{Re}(\omega)$ and the azimuthal wavenumber of the perturbation to the interface is $m$.
The linearized mass conservation (\ref{eq:nond1}a) equation gives
\begin{subequations}
\begin{equation} \label{eq:linMassCons} 
\sigma_j = -\frac{m\phi_j}{\omega r}\dv{\rho_j^*}{r}.
\end{equation}
The linearized radial equation of motion gives
\begin{multline} \label{eq:linRadial}
-\frac{m\omega\phi_j}{r} = -\frac{1}{\rho_j^*}\dv{P_j}{r} + \frac{\sigma_jr}{\rho_j^*} - 2\dv{\phi_j}{r} \\
 + \frac{\textrm{i}m}{\Rey}\frac{\mu_j^*}{\rho_j^*}\left\{\frac{1}{r}\dv{}{r}\left[r\dv{}{r}\left(\frac{\phi_j}{r}\right)\right]-\frac{(1+m^2)\phi_j}{r^3} + \frac{2}{r^2}\dv{\phi_j}{r} 
+\frac{2}{\mu_j^*}\dv{\mu_j^{*}}{r}\left[\frac{1}{r}\dv{\phi_j}{r} - \frac{\phi_j}{r^2}\right]
\right\},
\end{multline}
and the linearized azimuthal equation of motion is
\begin{multline} \label{eq:linAzimuthal} 
-\textrm{i}\omega\dv{\phi_j}{r} = -\frac{\textrm{i}mP_j}{\rho_j^* r} -2\frac{\textrm{i}m\phi_j}{r} \\
+ \frac{1}{\Rey}\frac{\mu_j^*}{\rho_j^*}\left\{-\frac{1}{r}\dv{}{r}\left(r\dv{^2\phi_j}{r^2}\right) 
+\frac{(1+m^2)}{r^2}\dv{\phi_j}{r} - \frac{2m^2\phi_j}{r^3} - \frac{1}{\mu_j^*}\dv{\mu_j^*}{r}\left[\dv{^2\phi_j}{r^2} - \frac{1}{r}\dv{\phi_j}{r} + \frac{m^2\phi_j}{r^2}\right]
\right\}.
\end{multline}
\end{subequations}
Eliminating $P_j$ and $\sigma_j$ from \eqref{eq:linRadial} using \eqref{eq:linMassCons} and \eqref{eq:linAzimuthal} yields a fourth order, one-dimesional linear ordinary differential equation, of an Orr-Sommerfeld type, for perturbations to the system given by
\begin{multline} \label{eq:os}
\textrm{i}\omega\left\{\left(\phi_j''+\frac{\phi_j'}{r}-\frac{m^2\phi_j}{r^2}\right) + \frac{\rho_j^{*}\,\!'}{\rho_j^*}\left(\phi_j'+\frac{m(m-2\omega)\phi_j}{\omega^2 r}\right)\right\}
 \\
 = \frac{1}{\Rey}\frac{\mu_j^*}{\rho_j^*}\left\{\phi_j''''+\frac{2\phi_j'''}{r}-(1+2m^2)\left[\frac{\phi_j''}{r^2}-\frac{\phi_j'}{r^3}\right]+\frac{m^2(m^2-4)\phi_j}{r^4}
 \right. \\ \left.
 + \frac{\mu_j^{*}\,\!'}{\mu_j^*}\left(2\phi_j'''+\frac{\phi_j''}{r}-\frac{(1+2m^2)\phi_j'}{r^2}+ \frac{3m^2\phi_j}{r^3}\right)
 +\frac{\mu_j^{*}\,\!''}{\mu_j^*}\left(\phi_j'' - \frac{\phi_j'}{r} + \frac{m^2\phi_j}{r^2}\right)
 \right\},
\end{multline}
where a prime symbol now indicates differentiation with respect to $r$.


\subsection{\label{sec:bcs}Boundary conditions}


In both the inviscid and viscous cases we require finite velocities on the axis $r=0$ and a no-penetration condition (no-normal-velocity condition) on $r=1$ such that
\begin{equation}
|\vc{u}(r=0)|<\infty,
\quad
\vc{u}(r=1)\cdot\hat{\vc{r}} = 0.
\end{equation}
In the case of a viscous outer fluid we will also require a no-slip condition on $r=1$, \textit{viz}.
\begin{equation}
\vc{u}(r=1)\cdot\hat{\vc{\theta}} = 0.
\end{equation}
These three conditions become, in terms of \eqref{eq:ueq}
\begin{equation} \label{eq:bcs1}
\lim_{r\to0}\left|\frac{\phi_1}{r}\right|<\infty, 
\quad
|\phi_1'(0)|<\infty,
\quad
\phi_2(1) = 0,
\quad
\phi_2'(1) = 0,
\end{equation}
where the first two conditions are the velocity regularity condition, the third is the no-penetration condition and the final condition is the no-slip condition that applies when the outer fluid is viscous.

The condition of stress continuity at the interface of two fluids with surface tension is given dimensionally by
\begin{equation}
\Delta\left\{\tn{\sigma}\cdot\hat{\vc{n}}\right\} = \gamma\left(\nabla\cdot\hat{\vc{n}}\right)\hat{\vc{n}}
\end{equation}
where $\Delta\left\{\cdot\right\}$ indicates the jump in a quantity from the outer fluid 2 to the inner fluid 1 across the interface $\mathscr{S}$, $\hat{\vc{n}}$ is a unit normal vector at the interface directed from fluid 1 into fluid 2, and $\gamma$ is the coefficient of surface tension.  Nondimensionalising (and dropping the prime notation immediately) we have
\begin{equation} \label{eq:stressCont}
\Delta\left\{\tn{\sigma}\cdot\hat{\vc{n}}\right\} = \frac{1}{\We}\left(\nabla\cdot\hat{\vc{n}}\right)\hat{\vc{n}},
\end{equation}
where $\We = \rho_0\Omega^2 a^3/\gamma$ is a nondimensional Weber number representing the ratio of inertial to curvature effects.

Taking the interface, $\mathscr{S}$, to be defined by
\begin{equation}
\mathscr{S}:=r - \left(r_0 + \epsilon\textrm{e}^{\textrm{i}(m\theta + \omega t)}\right) = 0, 
\end{equation}
then the unit vector pointing from fluid 1 into fluid 2 is given by
\begin{equation}
\hat{\vc{n}} = \frac{\nabla\mathscr{S}}{|\nabla\mathscr{S}|} = \left(1+O(\epsilon^2)\right)\hat{\vc{r}} +\left(- \frac{\textrm{i}m}{r}\epsilon\textrm{e}^{\textrm{i}(m\theta + \omega t)} + O(\epsilon^2)\right)\hat{\vc{\theta}}.
\end{equation}
Hence, on $\mathscr{S}$
\begin{gather} \label{eq:n}
\hat{\vc{n}} = \left(1+O(\epsilon^2)\right)\hat{\vc{r}} +\left(- \frac{\textrm{i}m}{r_0}\epsilon\textrm{e}^{\textrm{i}(m\theta + \omega t)} + O(\epsilon^2)\right)\hat{\vc{\theta}}, \\
\label{eq:divn}
\nabla\cdot\hat{\vc{n}} = \frac{1}{r_0} + \epsilon\frac{m^2 - 1}{r_0^2}\textrm{e}^{\textrm{i}(m\theta + \omega t)} + O(\epsilon^2).
\end{gather}

The stress tensor in each fluid layer is given in nondimensional terms by
\begin{equation} \label{eq:stressJump}
\tn{\sigma}_j = -p_j\tn{I} + \frac{2\mu_j}{\Rey}\tn{e}_j
\quad\Rightarrow
\quad
\Delta\left\{\tn{\sigma}\cdot\hat{\vc{n}}\right\} = \Delta\left\{-p\hat{\vc{n}}\right\} + \frac{2}{\Rey}\Delta\left\{\mu\tn{e}\cdot\hat{\vc{n}}\right\}.
\end{equation}
We consider the two terms on the right hand side separately.  Taylor expanding about $r=r_0$ we have that
\begin{equation}
\Delta\left\{-p\hat{\vc{n}}\right\} =- \left[p^* + \epsilon\left(\dv{p^*}{r}\Big|_{r = r_0} + P\right)\textrm{e}^{\textrm{i}(m\theta + \omega t)}
+O(\epsilon^2)\right]^+_-\,\hat{\vc{n}},
\end{equation}
where the jump on the right hand side is across $r=r_0$ (as distinct from the jump across the interface).  Substituting in both the hydrostatic condition $\textrm{d}p_j^*/\textrm{d}r = \rho_j^* r$ and the expression for $\hat{\vc{n}}$ on $\mathscr{S}$, \eqref{eq:n}, we may rewrite this last expression as
\begin{multline} \label{eq:pJump}
\Delta\left\{-p\vc{n}\right\} =  -\left\{\left[p^* + \epsilon\left(\rho^* r_0
+ P\right)\textrm{e}^{\textrm{i}(m\theta + \omega t)}\right]^+_- +O(\epsilon^2)\right\}\hat{\vc{r}} 
\\
+\left\{\frac{\epsilon\textrm{i}m}{r_0}\big[p^*\big]^+_-\,\textrm{e}^{\textrm{i}(m\theta + \omega t)}+O(\epsilon^2)\right\}\hat{\vc{\theta}}.
\end{multline}

For the flow under consideration, described in terms of \eqref{eq:ueq}, the rate of strain tensor $\tn{e}$ is given by
\begin{equation}
\tn{e} = \frac{\epsilon}{2r^2} \left(\begin{array}{cc}
2\,\textrm{i}\,m(r\phi' - \phi) & -r^2\phi'' + r\phi' - m^2\phi \\
-r^2\phi'' + r\phi' - m^2\phi & -2\,\textrm{i}\,m\left(r\phi' - \phi\right) 
\end{array}\right)\textrm{e}^{\textrm{i}(m\theta+\omega t)}.
\end{equation}
Hence, 
\begin{multline} \label{eq:eJump}
\Delta\left\{\mu\tn{e}\cdot\hat{\vc{n}}\right\} = \left\{\frac{\epsilon\textrm{i}m}{r_0^2}\big[\mu^*\left(r_0\phi' - \phi\right)\big]^+_-\,\textrm{e}^{\textrm{i}(m\theta+\omega t)}+O(\epsilon^2)\right\}\hat{\vc{r}} 
\\
+ \left\{\frac{\epsilon}{2r_0^2}\big[\mu^*\left(-r_0^2\phi'' + r_0\phi' - m^2\phi\right)\big]^+_-\,\textrm{e}^{\textrm{i}(m\theta+\omega t)}+O(\epsilon^2)\right\}\hat{\vc{\theta}}. 
\end{multline}
Combining \eqref{eq:n}, \eqref{eq:divn}, \eqref{eq:stressJump}, \eqref{eq:pJump} and \eqref{eq:eJump} with the stress continuity condition \eqref{eq:stressCont} we have at $O(1)$ and $O(\epsilon)$ in the normal direction and $O(\epsilon)$ in the tangential direction, respectively, the following conditions for stress continuity across $\mathscr{S}$
\begin{subequations} \label{eq:stressCont2}
\begin{equation}
\big[p^*\big]^+_- = -\frac{1}{\We}\frac{1}{r_0},
\end{equation}
\begin{equation}
\left[\rho^* r_0 + P   -\frac{2\mu^*}{\Rey}\frac{\textrm{i}m}{r_0^2}\left(r_0\phi' - \phi\right) \right]^+_-
= -\frac{1}{\We}\frac{m^2-1}{r_0^2},
\end{equation}
\begin{equation}
\big[\mu^*\left(r_0^2\phi'' - r_0\phi' + m^2\phi\right)\big]^+_-=0,
\end{equation}
\end{subequations}
where $(\ref{eq:stressCont2}a)$ has been used to simplify $(\ref{eq:stressCont2}c)$.

The kinematic condition at the interface, $\mathscr{S}$, is that the fluid at the interface should move with the velocity of the interface, hence
\begin{equation}
\Dv{}{t}\left(r - \left[ r_0 + \epsilon\textrm{e}^{\textrm{i}(m\theta+\omega t)}\right]\right)\Big|_{\mathscr{S}} = 0
\quad\Rightarrow\quad
u|_{\mathscr{S}}\sim\epsilon\textrm{i}\omega\textrm{e}^{\textrm{i}(m\theta+\omega t)}.
\end{equation}
The linearized kinematic condition in terms of \eqref{eq:ueq} gives
\begin{equation} \label{eq:twoLayerKin}
\phi_j(r_0) = \frac{\omega r_0}{m}, \quad j = 1,2.
\end{equation}
This condition may be also seen to match the normal fluid velocities at the interface.  In the case of two viscous fluids, the tangential fluid velocities at the interface are also forced to match, and this condition is satisfied at $O(\epsilon)$ when
\begin{equation} \label{eq:tanCont}
\phi_1'(r_0) = \phi_2'(r_0).
\end{equation}
Equations \eqref{eq:bcs1}, \eqref{eq:stressCont2}, \eqref{eq:twoLayerKin} and \eqref{eq:tanCont} are the complete set of boundary conditions for the problem.


\subsection{\label{sec:numSim}Numerical simulation}


A number of  numerical simulations were performed using a volume-of-fluid method.  The method was implemented using modifications of the `interFoam' and `twoLiquidMixingLayer' solvers, available as part of the OpenFOAM distribution \citep{openfoam}.  The standard implementation of the velocity equation was modified to account for the rotation by including both the centrifugal term and the Coriolis term, otherwise the solvers were unchanged.  An extensive review and discussion of the implementation of the interFoam solver is provided in \citet{deshpande_etal}.  The solver uses a volume-of-fluid approach to solve the continuity equation
\begin{equation}
\pd{\rho}{t} + \nabla\cdot\left(\rho\vc{u}\right) = 0,
\end{equation}
and momentum equation
\begin{multline} \label{eq:cfdmom}
\pd{}{t}\left(\rho\vc{u}\right) + \nabla\cdot\left(\rho\vc{u}\vc{u}\right) = -\nabla p + \left[\nabla\cdot\left(\mu\nabla\vc{u}\right) + \nabla\vc{u}\cdot\nabla\mu\right] 
\\ - \int_\varGamma \left[\gamma\nabla\cdot\hat{\vc{n}}\right] \delta\left(\vc{x} - \vc{x}_s\right)\hat{\vc{n}}\,\textrm{d}\varGamma(\vc{x}_s),
 \end{multline}
where $\gamma$ is the surface tension coefficient, $\delta$ is the Dirac delta function in this instance, and $\varGamma$ denotes the interface between the two phases.  The momentum equation was modified by including the rotational terms, $ - \rho\,\vc{\Omega}\times\left(\vc{\Omega}\times\vc{x}\right) - 2\rho\,\vc{\Omega}\times\vc{u}$ on the right hand side of \eqref{eq:cfdmom}.  A PISO \citep{issa} based predictor-corrector method is then used to first predict an updated velocity field and then correct and update the pressure and velocity fields, enforcing incompressibility.  The circular domain was meshed with a 5-block mesh and typically contained $4.8\times10^5$ cells.  Time was stepped forward using a Crank-Nicolson scheme, limited by the CFL number. 


\section{\label{sec:theComp}Theoretical predictions and comparison with numerics}



\subsection{\label{sec:crti}Two-layer stable and unstable inviscid solutions: Rayleigh approximation}


The simplest configuration we consider consists of two layers of inviscid fluid that have a density difference.  In this case $\mu_j^*(r)=0$ and $\rho_j^*(r)=\rho_j$ is constant in each layer such that $\textrm{d}\rho_j^*/\textrm{d}r = 0$.  The initial hydrostatic conditions for the velocity and density are
\begin{equation} \label{eq:initStrat}
\vc{u}_j^* = \vc{0},
\quad
\rho_j^* = \left\{\begin{array}{ll}
\rho_1 & r < r_0, \\
\rho_2 & r > r_0,
\end{array}\right.
\quad
p_j^* = p_0 + \left\{\begin{array}{ll}
\rho_1 r^2/2 & r < r_0, \\
(\rho_1-\rho_2) r_0^2/2 + \rho_2 r^2/2 & r > r_0
\end{array}\right.
\end{equation}
where $r = r_0\in(0,1)$ is the location of the initial unperturbed interface and $p_0$ is a reference pressure at the origin.  The $O(1)$ pressure continuity condition (\ref{eq:stressCont2}a) has been applied.  In the special case of equal volumes of fluid in each layer $r_0 = 2^{-1/2}$.

The form of $\rho_j^*$ and $\mu_j^*$ leads to the simplification of \eqref{eq:os}, removing all the viscous effects leaving a one-dimensional Laplace equation that may be considered a Rayleigh equation.  Specifically, we define the linear differential operator $\mathcal{L}$ and have
\begin{equation} \label{eq:rayleighapprox}
\mathcal{L}[\phi] := \phi_j''+\frac{\phi_j'}{r}-\frac{m^2\phi_j}{r^2}=0,
\end{equation}
for $\omega\ne0$, $j=1, 2$, together with the pressure perturbation
\begin{equation}
P_j = -\rho_j\left(2\phi_j - \frac{\omega r}{m}\phi_j'\right).
\end{equation}
Equation \eqref{eq:rayleighapprox} yields power-law solutions for $m\geqslant1$ given by
\begin{equation}
\phi_j = c_{j1}r^{-m} + c_{j2}r^{m}, \quad j=1,2,
\end{equation}
with four unknown constants.  The dispersion relation may now be found by fitting appropriate matching and boundary conditions.  

We enforce the velocity regularity condition on $r=0$ and the no-penetration condition on $r=1$.  We then apply the kinematic condition and the stress continuity condition on the perturbed interface $r = r_0 + \epsilon\exp\{\textrm{i}(m\theta+\omega t)\}$ as described in \S\,\ref{sec:bcs}.  These five conditions with only four free constants then lead to the dispersion relation.  The velocity regularity condition on $r=0$ implies that $|\phi_1/r|<\infty$ as $r\to0$ and so we take $c_{11}=0$.  The no-penetration condition on $r=1$ implies that $\phi_2(1)=0$ and so we take $c_{22} = -c_{21}$. 

Applying the kinematic condition \eqref{eq:twoLayerKin} then gives
\begin{equation} \label{eq:phisol}
\phi_1 = \frac{\omega r_0}{m}\left(\frac{r}{r_0}\right)^m,
\quad
\phi_2 = \frac{\omega r_0}{m}\left(\frac{r}{r_0}\right)^m\left(\frac{1-r^{-2m}}{1-r_0^{-2m}}\right).
\end{equation}

The continuity of stress at the interface now gives the dispersion relation.  In the present inviscid case, in the absence of surface tension, we have continuity of pressure across $\mathscr{S}$.  The tangential stress condition (\ref{eq:stressCont2}c) is automatically satisfied as $\mu_j^*=0$, and the leading order pressure continuity condition (\ref{eq:stressCont2}a) is satisfied by the hydrostatic solution \eqref{eq:initStrat}.  The remaining $O(\epsilon)$ pressure continuity condition, (\ref{eq:stressCont2}b), simplifies to
\begin{equation}
\rho_1\left[r_0 - 2\phi_1(r_0) + \frac{\omega r_0}{m}\phi_1'(r_0)\right] = \rho_2\left[r_0 - 2\phi_2(r_0) + \frac{\omega r_0}{m}\phi_2'(r_0)\right].
\end{equation}
Subsituting in \eqref{eq:phisol} leads to the dispersion relation
\begin{equation} \label{eq:disprel1}
\mathscr{A}\left(\omega^2 - 2\omega + m\right) = \frac{\left(1+\mathscr{A}\right)\omega^2}{1-r_0^{2m}},
\end{equation}
where we have used an Atwood number, $\mathscr{A}$, as in \citet{prf}, defined by
\begin{equation}
\mathscr{A} = \frac{\rho_2 - \rho_1}{\rho_2 + \rho_1}\in[-1, 1].
\end{equation}
We can further simplify \eqref{eq:disprel1} by introducing a modified Atwood number, $\mathscr{A}^*$, given by
\begin{equation}
\mathscr{A}^* = \mathscr{A}\left(\frac{1 - r_0^{2m}}{1 + \mathscr{A}r_0^{2m}}\right) \in [-1, 1],
\end{equation}
that has the same sign as $\mathscr{A}$ but has a shape factor based on the initial position of the interface, the azimuthal wavenumber and the Atwood number.  In terms of the modified Atwood number we may rewrite \eqref{eq:disprel1} as
\begin{equation} \label{eq:disprel2}
\omega^2 + \mathscr{A}^*\left(2\omega - m\right) = 0.
\end{equation}

In the special case $\mathscr{A}=\mathscr{A}^*=0$, that occurs when there is no density contrast, we can immediately show that $\omega=0$ and there is no growth or precession of the perturbation.

For the special case $\mathscr{A}=\mathscr{A}^*=-1$, that occurs when there is no fluid in the outer layer, we find $\omega = 1 \pm \textrm{i}\sqrt{m-1}$, i.e., except for $m=1$ all modes grow in amplitude.  When $m=1$ the inner fluid layer, the only fluid layer present, remains circular, but is displaced, precessing about the origin.  Note that $m=1$ is the only perturbation mode that gives rise to a non-zero fluid velocity at the origin of the system and may be considered a centre of mass oscillation.

Since we choose the form of the interface to be $r = r_0 + \epsilon\exp\{\textrm{i}(m\theta+\omega t)\}$, a negative imaginary part in $\omega$ corresponds to growth of the interface.  The dispersion relation \eqref{eq:disprel1} has solutions with negative imaginary parts when the discriminant of \eqref{eq:disprel1} is negative.  This condition is met when $\mathscr{A}^*<0$, i.e., $\mathscr{A}\in[-1,0)$ (except for the special case $m=1$, $\mathscr{A}=-1$ discussed above.):  we have unstable growth when the Atwood number is negative, i.e., when the density of the fluid in the inner layer is greater than that in the outer layer, as may be anticipated.

The solution to \eqref{eq:disprel2} is
\begin{equation} \label{eq:invSol}
\omega = -\mathscr{A}^* \pm \left\{\mathscr{A}^*\left(\mathscr{A}^* + m\right)\right\}^{1/2}
\end{equation}
When the Atwood number is negative then perturbations initially grow exponentially in time with precessional rate $-\mathscr{A}^*$ and growth rate $\left\{-\mathscr{A}^*\left(\mathscr{A}^* + m\right)\right\}^{1/2}$.  In the limit of large azimuthal wavenumber, $m\to\infty$, then $r_0^{2m}\to0$ and
\begin{equation} \label{eq:largeMLimit}
\omega\sim -\mathscr{A}\pm\left\{\mathscr{A}\left(\mathscr{A}+m\right)\right\}^{1/2},
\end{equation}
independently of $r_0$, as the curvature of the interface is not `felt' by the system for the high azimuthal wavenumbers as the azimuthal wavelength is too short compared to $r_0$.  The stable inviscid dispersion relation, \eqref{eq:disprel1} with $\mathscr{A}>0$, necessarily has one positive root and one negative root, denoted $\omega_\infty^\pm$, as a result of $m\left(1+\mathscr{A}r_0^{2m}\right)>0$.  One wave precesses clockwise, while the other precesses in an anticlockwise direction.

We can gain further insight into the physical mechanisms at work by proceeding dimensionally.  Distinguishing between the centrifugal term in the governing equation of motion \eqref{eq:Mom1}, $\vc{\Omega}\times\left(\vc{\Omega}\times\vc{x}\right)$, and the Coriolis term, $2\hat{\vc{\Omega}}\times\vc{u}$, by introducing a `hat' on the rotation vector in the Coriolis term, the dimensional dispersion relation \eqref{eq:disprel2} is given by
\begin{equation} \label{eq:dimDisp}
\omega^2 + \mathscr{A}^*\left(2\hat{\Omega}\omega - \Omega^2m\right) = 0,
\end{equation}
where $\mathscr{A}^* = \mathscr{A}\left(a^{2m} - r_0^{2m}\right)/\left(a^{2m} + \mathscr{A}r_0^{2m}\right)$ and $\hat{\Omega}\equiv\Omega$: the hat notation is used only to highlight terms originating from Coriolis forces.  The constant term in \eqref{eq:dimDisp} is now identifiable with the centrifugal term from the equation of motion \eqref{eq:Mom1}, while the linear term is associated with the Coriolis term.  The discriminant of \eqref{eq:dimDisp} is proportional to $\mathscr{A}^{*2}\hat{\Omega}^2 + \mathscr{A}^*\Omega^2m$ where the first term can now be associated with Coriolis effects and the second term with centrifugal effects.  In order for a perturbation to grow in time the discriminant must be negative.  However, the first term of the discriminant is always positive and so Coriolis effects can never lead to unstable growth.   Any growth that occurs is due to the second term of the discriminant which is associated with the centrifugal term in \eqref{eq:Mom1} and hence any growth is therefore `centrifugally driven'.  For any configuration where we have centrifugally driven growth, the Coriolis term must slow the growth rate but can never prevent growth as a result of $|\mathscr{A}^{*2}\hat{\Omega}^2| < |\mathscr{A}^*\Omega^2m|$.

The dispersion relation of \citet{tao} is recovered in the limit of: a hydrostatic initial interface, $g = R\Omega^2$ (in their notation); and $-kR = m$ where $m$ is our azimuthal wavenumber and $k$ is their Cartesian wavenumber.  The minus sign accounts for the difference in the sense of our $\hat{\vc{\theta}}$ and their $\hat{\vc{x}}$.  Finally $\mathscr{A}^*$ is replaced by $-\mathscr{A}$, due to neglecting the curvature of the interface and taking an unbounded outer domain ($r_0/a\to 0$), the sign change is due to the difference in definition of the Atwood number between the two treatments.


\subsection{\label{sec:surfaceTension}Surface tension between immiscible fluids}


Returning to nondimensional quantities, we now consider the effect of surface tension acting between two inviscid immiscible fluid layers.  The pressure field in the hydrostatic solution is modified from the previous case considered in \S\,\ref{sec:crti}, due to the finite Weber number in (\ref{eq:stressCont2}a), and is now given by
\begin{equation}
p_j^* = p_0 + \left\{\begin{array}{ll}
 \rho_1 r^2/2 + 1/(\We\, r_0) & r < r_0, \\
(\rho_1-\rho_2) r_0^2/2 + \rho_2 r^2/2 & r > r_0
\end{array}\right.
\end{equation}
while the hydrostatic, velocity and density fields remain as before.  The kinematic conditions are as in \eqref{eq:twoLayerKin}, hence \eqref{eq:phisol} remains unchanged.  The $O(\epsilon)$ normal stress continuity condition at the interface (\ref{eq:stressCont2}b) is
\begin{equation} \label{eq:stJump}
\rho_1\left(r_0-2\phi_1(r_0) + \frac{\omega r_0}{m}\phi_1'(r_0)\right)
 = \rho_2\left(r_0 - 2\phi_2(r_0) + \frac{\omega r_0}{m}\phi_2'(r_0)\right)+\frac{1}{\We}\frac{m^2-1}{r_0^2}.
\end{equation}
This modifies \eqref{eq:disprel2} to give
\begin{equation} \label{eq:disprelST}
\omega^2 + \mathscr{A}^*\left(2\omega - m\right) = \frac{\mathscr{A}^*}{\We}\frac{m\left(m^2-1\right)}{2\mathscr{A}r_0^3}=: \mathscr{A}^* S(\We;\mathscr{A}, m, r_0),
\end{equation}
defining the surface tension adjustment factor $S$.  It follows immediately from the form of \eqref{eq:disprelST} that surface tension has the greatest effect when the wavenumber, $m$, is high and when the mean radius of the interface, $r_0$, is small, as expected on physical grounds.  The result \eqref{eq:disprelST} agrees with the results for a rotating column of liquid \citep{hockingMichael} and a rotating inviscid drop \citep{patzeketal}, up to a change of frame of reference and nondimensionalisation, in the special case $\mathscr{A}=-1$.

The solution to \eqref{eq:disprelST}, modified compared to \eqref{eq:invSol}, is
\begin{equation} \label{eq:dispRelST2}
\omega = -\mathscr{A}^* \pm \left\{\mathscr{A}^*\left(\mathscr{A}^* + m + S\right)\right\}^{1/2}.
\end{equation}
Inspection of \eqref{eq:dispRelST2} and comparison with \eqref{eq:invSol} shows that the effect of the surface tension can be interpreted as modifying the azimuthal wave number.  In a stable configuration the effect is to increase the apparent azimuthal wavenumber, as $S>0$, enhancing the frequency of oscillation, as might be anticipated since the surface tension applies an additional restorative force on the interface.  In an unstable configuration the effect is to decrease the apparent azimuthal wavenumber, as $S<0$, inhibiting the growth of the instability.  The surface tension affects the constant term in the quadratic dispersion relation \eqref{eq:dispRelST2} and so influences the contribution of the centrifugal forcing to the system.  It is therefore possible for the surface tension not only to influence the growth rate of unstable modes but indeed completely stabilize an otherwise unstable mode if the surface tension is large enough.  It follows from \eqref{eq:dispRelST2} that for a naturally unstable mode, $m$, the surface tension is able to completely stabilize the mode when $S\leqslant -\left(\mathscr{A}^* + m\right)$.  Alternatively, we observe that the surface tension is able to completely suppress the growth of high frequency modes above a critical value $m^*$ given by the implicit relationship $-m^* = \mathscr{A}^* + S(\We;\mathscr{A},m^*, r_0)$ or 
\begin{equation} \label{eq:surfTenCutOff}
\mathscr{A}(1-r_0^{2m^*}) + (1+\mathscr{A}r_0^{2m^*})m^*\left[1+\frac{1}{\We}\frac{(m^{*2}-1)}{2\mathscr{A}r_0^3}\right] = 0.
\end{equation}

When $r_0$ is close to the outer boundary, say $r_0 = 1 - \varepsilon$, $\varepsilon\ll1$, [n.b.\, $\varepsilon$ is different to the small parameter $\epsilon$ in \eqref{eq:pertEqs}, it is assumed that $0\ll\epsilon\ll\varepsilon\ll1$] then we find 
\begin{equation}
\omega\sim m\left\{\pm\left(1 + \frac{1}{\We} \frac{m^2-1}{2\mathscr{A}}\right)^{1/2}\left(\frac{2\mathscr{A}}{1+\mathscr{A}}\varepsilon\right)^{1/2}-\frac{2\mathscr{A}}{1+\mathscr{A}}\varepsilon + O(\varepsilon^{3/2})\right\},
\end{equation}
and we see that a given mode with wavenumber $m$ is stable for $\We > -(m^2-1)/2\mathscr{A}$ (cf.~\eqref{eq:surfTenCutOff} as $r_0\to1$).


\subsection{\label{sec:diff}Diffusion of the interface between miscible fluids}

\begin{figure} 
\begin{center}
\hspace*{-20pt}\includegraphics[clip = TRUE]{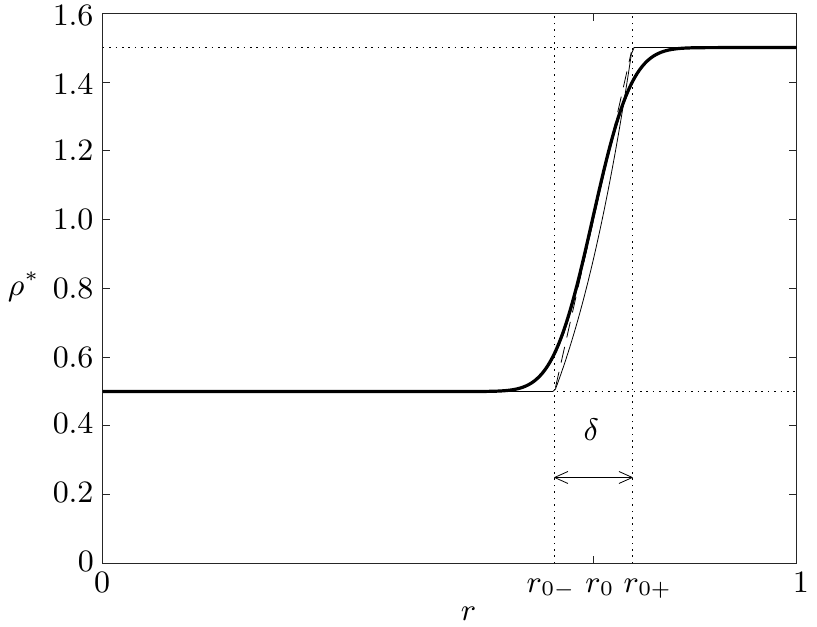}
\end{center}
\caption{\label{fig:deltaApprox}A comparison of the approximations to the diffuse layer for $\mathscr{A} = \frac{1}{2}$, $t/\textrm{Pe} = 10^{-3}$, $r_0 = 2^{-1/2}$.  The exact solution \eqref{eq:diffsol} is shown as solid bold.  The similarity solution \eqref{eq:simSol} is plotted but cannot be distinguished by eye at this scale from the exact solution.  The piecewise power-law approximation \eqref{eq:densityStrat} is shown as a solid line.  The piecewise linear approximation is shown dashed for comparison.  Dotted lines indicate the width of the diffuse layer $\delta$, given in \eqref{eq:deltaApprox}.}
\end{figure}

It is well-known in the study of Kelvin-Helmholtz instability that diffusion of the interface between the two fluid layers can lead to some modes of instability being suppressed.  Here we investigate whether a diffuse fluid layer between the inner and outer fluids causes some modes of centrifugally driven Rayleigh-Taylor instability to be suppressed similarly.  We begin by considering the form of a diffuse interface between the two layers.  We consider a two-layer stratification with a sharp interface (as in \eqref{eq:initStrat}, with $\rho^*=\rho_1$ in the inner layer and $\rho^*=\rho_2$ in the outer layer) that is subject to diffusion of density governed by the dimensional diffusion equation
\begin{equation} \label{eq:govDif}
\Dv{\rho^*}{t} = \kappa\nabla^2\rho^*,
\end{equation}
where $\kappa$ is a constant diffusivity.  A natural scaling for this problem in isolation is to take $t = a^2\kappa^{-1}t'$, however to enable comparison with the temporal development of the instability we choose $t = \Omega^{-1}t'$. Therefore, nondimensionalising as in \S\,\ref{sec:model} and dropping the prime notation we obtain
\begin{equation}
\Dv{\rho^*}{t} = \frac{1}{\Pe}\nabla^2\rho^*,
\end{equation}
where $\Pe = \Omega a^2/\kappa$ is a nondimensional P\'eclet number that relates the advection timescale to the diffusion timescale.  We consider the diffusion of density in the hydrostatic fluid such that $\rho^* = \rho^*(r, t)$, $\vc{u}^*=\vc{0}$ and hence
\begin{equation} \label{eq:diffeq}
\pd{\rho^*}{t} = \frac{1}{\Pe} \frac{1}{r}\pd{}{r}\left(r\pd{\rho^*}{r}\right),
\quad r\in(0,1),~ t\in[0,\infty)
\end{equation}
The initial condition is as in \eqref{eq:initStrat} and we enforce Neumann boundary conditions $\partial\rho^*/\partial r = 0$ on $r=0, 1$.  By separation of variables a series solution to the above problem can be found and is given by
\begin{equation} \label{eq:diffsol}
\frac{\rho^*}{\rho_0} = 1 + \mathscr{A} - 2\mathscr{A}r_0 \left\{r_0 + 2\sum_{n = 1}^\infty \frac{\mathcal{J}_1(\lambda_n r_0)\mathcal{J}_0(\lambda_n r)}{\lambda_n\mathcal{J}_0(\lambda_n)^2}\textrm{e}^{-\lambda_n^2t/\Pe}\right\},
\end{equation}
where $\mathcal{J}$ is a Bessel function of the first kind and $\lambda_n$ for $n=1,2,\ldots$ are the zeros of $\mathcal{J}_1$, i.e., $\lambda_1 \approx 3.83$, $\lambda_2 \approx 7.02$ etc.  It can be seen that in the long-time limit the density is constant everywhere equal to the mean density.  The `thickness' of the diffuse layer, to be denoted by $\delta$, is subjective, but one method is to define the thickness by
\begin{equation} \label{eq:layerThickness}
\delta = \left(\rho_2 - \rho_1\right)\left(\left.\pd{\rho^*}{r}\right|_{r = r_0}\right)^{-1},
\end{equation}
that follows from fitting a piecewise linear density profile that has the same gradient as the full solution at $r=r_0$.  An approximate scaling for $\delta$ can be found as follows.  If we consider the diffuse layer to be small, such that the rapid changes in $\rho^*$ take place in a narrow region $r = r_0 + \varepsilon x$, where $\varepsilon\ll1$ (and again, $\varepsilon$ is an arbitrary small parameter, not related to $\epsilon$ in \S\,\ref{sec:model}) and the time scale for diffusion is taken such that $t = \varepsilon^2 \Pe\, \tau$ then \eqref{eq:diffeq} becomes
\begin{equation}
\pd{\rho^*}{\tau} = \pd{^2\rho^*}{x^2} + O(\varepsilon),
\end{equation}
which may therefore be considered Cartesian at leading order as the curvature effects are negligible.  This form of the diffusion equation accepts a similarity solution
\begin{equation}
\rho^* = \frac{\rho_1+\rho_2}{2} + \frac{\left(\rho_2 - \rho_1\right)}{2}\textrm{erf}\left(\frac{x}{2\sqrt{\tau}}\right),
\end{equation}
where $\textrm{erf}(x)$ is the error function $2\pi^{-1/2}\int_0^x \exp\{-\xi^2\}\,\textrm{d}\xi$ and $\rho^*$ satisfies the boundary conditions $\rho^*\to\rho_1$ as $x\to-\infty$, $\rho^*\to\rho_2$ as $x\to\infty$, and the initial conditions \eqref{eq:initStrat}.  Therefore, at early times we may approximate the diffuse layer given by \eqref{eq:diffsol} as
\begin{equation} \label{eq:simSol}
\frac{\rho^*}{\rho_0} = 1 +\mathscr{A}\textrm{erf}\left(\frac{r - r_0}{2\sqrt{t/\Pe}}\right).
\end{equation}
Figure \ref{fig:deltaApprox} shows the solution of \eqref{eq:govDif} as given in \eqref{eq:diffsol} (thick solid line) compared to the early time approximation \eqref{eq:simSol} (thin solid line, indistinguishable from the full solution at the scale shown).  The form of solution in \eqref{eq:simSol} leads to a simple approximation for $\delta$, as defined in \eqref{eq:layerThickness}, giving
\begin{equation} \label{eq:deltaApprox}
\delta \sim 2\sqrt{\pi t/\Pe} + O(\varepsilon),
\end{equation}
which is also shown in figure \ref{fig:deltaApprox}.

We consider the effect of diffusion on a system with large P\'eclet number, where diffusion is slow compared to the growth of an instability, or the oscillation of an interfacial wave.  This would correspond to an experimental situation where the interface had been allowed to diffuse before the experiment was conducted, but during the experiment the effects of further diffusion could be neglected.  As such we seek to approximate the effect of diffusion by constructing a piecewise density profile.  Examining \eqref{eq:os} we see that if we can approximate the diffuse interface by a piecewise density $\rho^*$ such that $\rho^*$ is of the form $\rho^* = \beta r^{\alpha}$ for some constants $\beta$ and $\alpha$, then the form of \eqref{eq:os} is unchanged, and power-law solutions are still admitted.  We define $r_{0-} = r_0-\delta/2$, the inner edge of the diffuse layer, and $r_{0+} = r_0 + \delta/2$, the outer edge of the diffuse layer and identify a three-layer system $(j\in\{1, 2, 3\})$ where
\begin{equation} \label{eq:densityStrat}
\rho^* = \left\{\begin{array}{lrcll}
\rho_1 & 0 & \!\!\! \leqslant r < & \!\!\! r_{0-}; & j=1 \\
\beta r^\alpha & r_{0-} & \!\!\! \leqslant r < & \!\!\! r_{0+}; & j = 2 \\
\rho_2 & r_{0+} & \!\!\! \leqslant r \leqslant &  \!\!\! 1; & j=3
\end{array}\right.
\end{equation}
and where $\alpha$ and $\beta$ are chosen to make the density continuous and are given by
\begin{equation}
\alpha = \log\left(\frac{1-\mathscr{A}}{1+\mathscr{A}}\right)\left[\log\left(\frac{r_{0-}}{r_{0+}}\right)\right]^{-1},
\quad
\beta =\frac{1-\mathscr{A}}{r_{0-}^\alpha}.
\end{equation}
It follows that
\begin{equation}
\frac{\rho^*\,\!'}{\rho^*} = \left\{\begin{array}{lrcl}
0 & 0 & \!\!\! \leqslant r < & \!\!\! r_{0-} \\
\alpha/r & r_{0-} & \!\!\! \leqslant r < & \!\!\! r_{0+} \\
0 & r_{0+} & \!\!\! \leqslant r \leqslant &  \!\!\! 1
\end{array}\right.
\end{equation}
The corresponding initial hydrostatic pressure field is
\begin{equation}
p^* = p_0 + \left\{\begin{array}{lrcl}
\displaystyle\frac{\rho_1}{2}r^2 & 0 & \!\!\! \leqslant r < & \!\!\! r_{0-} \\[2ex]
\displaystyle\frac{\rho_1}{2}r_{0-}^2 + \frac{\beta}{2+\alpha}\left[r^{2+\alpha} - r_{0-}^{2+\alpha}\right]  & r_{0-} & \!\!\! \leqslant r < & \!\!\! r_{0+} \\[2ex]
\displaystyle\frac{\rho_1}{2}r_{0-}^2 
+ \frac{\beta}{2+\alpha}\left[r_{0+}^{2+\alpha} - r_{0-}^{2+\alpha}\right]
 -\frac{\rho_2}{2}r_{0+}^2 + \frac{\rho_2}{2}r^2
& r_{0+} & \!\!\! \leqslant r \leqslant &  \!\!\! 1
\end{array}\right.
\end{equation}

The governing equation \eqref{eq:os} is as in the two-layer no-diffusion case in the inner ($j=1$) and outer ($j=3$) layers and so we have
\begin{equation}
\phi_1(r) = c_{11} r^{-m} + c_{12}r^m, \quad
\phi_3(r) = c_{31} r^{-m} + c_{32}r^m,
\end{equation}
The governing equation \eqref{eq:os} in the transitional ($j=2$) layer is
\begin{equation}
\left(\phi_2''+\frac{\phi_2'}{r}-\frac{m^2\phi_2}{r^2}\right) + \frac{\alpha}{r}\left(\phi_2'+\frac{m(m-2\omega)\phi_2}{\omega^2 r}\right) = 0.
\end{equation}
This yields
\begin{equation}
\phi_2''+(1+\alpha)\frac{\phi_2'}{r} - \left(m^2 - \frac{m(m-2\omega)\alpha}{\omega^2 }\right)\frac{\phi_2}{r^2} = 0,
\end{equation}
with the power-law solution
\begin{equation} \label{eq:gammadef}
\phi_2(r) = c_{21}r^{-\alpha/2 - \chi} + c_{22}r^{-\alpha/2 + \chi}
,\quad\textnormal{where}\quad
\chi = \frac{1}{2}\left\{\alpha^2 + 4\left(m^2-\frac{m(m-2\omega)\alpha}{\omega^2 }\right)\right\}^{1/2}.
\end{equation}

We apply the velocity regularity condition at $r=0$ and no-penetration condition at $r=1$ to find $c_{11}=0$ and $c_{32} = -c_{31}$.
The boundary conditions are as derived in \S\,\ref{sec:bcs}, but are applied at $r_{0-}$ and $r_{0+}$.  These may be shown to be that $\phi_1 = \phi_2$ and $\phi_1' = \phi_2'$ at $r=r_{0-}$, and $\phi_2 = \phi_3$ and $\phi_2' = \phi_3'$ at $r=r_{0+}$.  Enforcing these matching conditions, to write $c_{12}$, $c_{22}$ at $r_{0-}$ and $c_{22}$ and $c_{31}$ at $r_{0+}$ in terms of $c_{12}$ leaves two expressions for the ratio $c_{21}/c_{22}$ which, combined, yield the dispersion relation
\begin{equation} \label{eq:diffdr}
\frac{2(\chi+m)+\alpha}{2(\chi-m)-\alpha}
\left[\frac{(2(\chi-m)-\alpha)r_{0+}^{2m}-2(\chi+m)+\alpha}{(2(\chi+m)+\alpha)r_{0+}^{2m}-2(\chi-m)-\alpha} \right]
= \left(\frac{r_{0-}}{r_{0+}}\right)^{2\chi}.
\end{equation}
For small diffuse layers where $\delta\ll1$, and $\delta$ may be approximated using \eqref{eq:deltaApprox} we have from \eqref{eq:diffdr}, that
\begin{multline}
\chi^2 \sim \left(\frac{\alpha_0 r_0}{2\delta}\right)^2\left\{1+\frac{4m}{\mathscr{A}^* \alpha_0 r_0}\delta \right. \\ \left.
-\left[\frac{1}{6r_0^2}-\left(\frac{2m}{\alpha_0 r_0}\right)^2\left(1+\frac{1+\mathscr{A}}{\mathscr{A}^3}\frac{(2\mathscr{A}+\alpha_0)[1 - \mathscr{A}(1-r_0^{2m})]}{(1-r_0^{2m})^2}\right)\right]\delta^2+O(\delta^3)\right\},
\end{multline}
where $\alpha_0 = \log[(1-\mathscr{A})/(1+\mathscr{A})]$.  It follows that
\begin{equation}
\omega\sim \omega_\infty\left\{1 - \left(\frac{m}{\alpha_0r_0}\right)\left(\frac{\mathscr{A}^*\omega_\infty}{\mathscr{A}^* + \omega_\infty}\right)
\frac{1+\mathscr{A}}{2\mathscr{A}^3}\frac{(2\mathscr{A}+\alpha_0)[1 - \mathscr{A}(1 - r_0^{2m})]}{(1 - r_0^{2m})^2}\delta + O(\delta^2)\right\},
\end{equation}
where $\omega_\infty$ is the zero-diffusion solution from \eqref{eq:invSol}.  It follows immediately that as the thickness of the diffuse layer, $\delta$, tends to zero, the zero-diffusion solution $\omega_\infty$ is recovered.

\begin{figure}
\begin{center}
\hspace*{-20pt}\includegraphics[clip = TRUE]{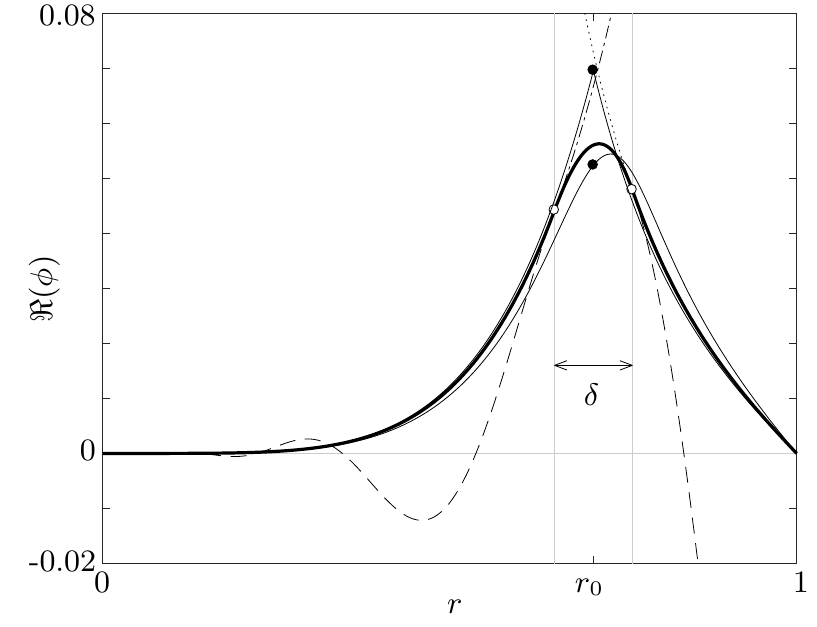}
\end{center}
\caption{\label{fig:diffComp}Comparison of the effect of the various models of the diffuse interface on $\phi$.  The no-diffusion case is the solid piecewise line comprising two solid thin line segments that are joined at the black circle.  The piecewise diffuse case is the thick black line, the solution is in three segments that coincide at the white circles.  $\phi_1$ continues dot-dashed after the first white circle, $\phi_2$ outside the diffuse region is shown dashed, and $\phi_3$ is shown dotted to the left of the second white circle.  The numerical solutions corresponding to the exact solution of the diffusion equation and the similarity solution of the diffusion equation are shown as a thin solid line and a dashed line, but the difference cannot be seen by eye at this scale.  The parameters chosen were: $\mathscr{A} = -\frac{1}{2}$, $t = 10^{-3}\Pe$, $m = 5$, $r_0 = 2^{-1/2}$.  The corresponding values of $\omega$ were: for the non-diffuse case $\omega=0.49 - 1.49\textrm{i}$; for the piecewise diffusion solution $\omega = 0.40 - 1.35\textrm{i}$; for the exact and similarity solution diffusion solutions $\omega = 0.37 - 1.31\textrm{i}$.}
\end{figure}

Figure \ref{fig:diffComp} compares the solutions $\phi$ for the various approximations to the diffuse layer with the sharp zero-diffusion solutions.  The zero-diffusion solution \eqref{eq:phisol} is shown as a thin solid line where the transition between $\phi_1$ and $\phi_2$ takes place at the solid black circle at a cusp.  For the chosen parameters, $\mathscr{A}=-\frac{1}{2}$, $m=5$, $r_0 = 2^{-1/2}$, the dispersion relation \eqref{eq:disprel1} gives $\omega = 0.49 - 1.49\textrm{i}$ for the most unstable mode.  We then considered the change in $\omega$ if the interface was allowed to diffuse over a time $t=10^{-3}\Pe$ prior to any perturbation to the system.  We can use the exact solution for $\rho^*$ \eqref{eq:diffsol} and then numerically solve the appropriate form of \eqref{eq:os} given by
\begin{equation} \label{eq:invdiff}
\left(\phi'' + \frac{\phi'}{r} - \frac{m^2\phi}{r^2}\right) + \frac{\rho^*\,\!'}{\rho^*}\left(\phi' + \frac{m(m-2\omega)\phi}{\omega^2r}\right)=0,
\end{equation}
across the whole domain to find $\phi$ and $\omega$.  This solution is shown as a thin solid line, where the kinematic condition has been enforced at $r = r_0$ indicated by the lower black circle.   The solution to \eqref{eq:invdiff} subject to $\phi(0) = 0$, $\phi(1) = 0$ was calculated separately in $[0, r_0]$ and $[r_0, 1]$ with the kinematic condition enforced for each solution at $r=r_0$.  The eigenvalue $\omega$ is chosen to ensure that $\phi$ is at least class $C^2[0,1]$.  The method of solution was to formulate the problem in terms of Chebyshev polynomials (see \citet{driscoll} for details) and the method generalizes well to the fully viscous case, as is discussed in \S\,\ref{sec:dvl}.  The same method can be implemented using the similarity solution \eqref{eq:simSol}, and this is shown as a dashed line, though it is hard to distinguish from the exact solution at this scale.  These numerical solutions yield an eigenvalue $\omega =0.37-1.31\textrm{i}$ for the most unstable mode, showing that the diffusion inhibits the most unstable mode.  Finally, the piecewise solution is shown as the thick bold line and comprises three segments that transition at the white circles located on $r=r_{0\pm}$.  The $\phi_1$ solution is plotted, dot-dashed, to the right of the left-hand white circle though it is no-longer part of the solution.  Similarly, $\phi_2$ is shown to the left of $r_{0-}$ and the right of $r_{0+}$ dashed.  The final part of the solution, $\phi_3$, is shown to the left of the right-hand white circle, dotted, before it becomes part of the solution.  The predicted most unstable mode in this case has eigenvalue $\omega = 0.40 - 1.35\textrm{i}$.


\subsection{\label{sec:vfl}Viscous fluid layers}


We now consider the case of two fluid layers separated by a sharp interface with differing, but constant, densities and viscosities.  We begin by considering the fluids to be miscible with no surface tension acting at the interface.  In each fluid layer $j=1, 2$, the governing equation \eqref{eq:os} simplifies as
\begin{multline} \label{eq:osVisc}
\textrm{i}\omega\left\{\left(\phi_j''+\frac{\phi_j'}{r}-\frac{m^2\phi_j}{r^2}\right)\right\}
 \\
 = \frac{1}{\Rey}\frac{\mu_j^*}{\rho_j^*}\left\{\phi_j''''+\frac{2\phi_j'''}{r}-(1+2m^2)\left[\frac{\phi_j''}{r^2}-\frac{\phi_j'}{r^3}\right]+\frac{m^2(m^2-4)\phi_j}{r^4}\right\}.
\end{multline}
This can be rewritten in terms of the differential operator, $\mathcal{L}$, defined in \eqref{eq:rayleighapprox} as
\begin{equation}
\textrm{i}\omega \mathcal{L}[\phi] = \frac{1}{\Rey}\frac{\mu_j^*}{\rho_j^*}\mathcal{L}^2[\phi].
\end{equation}
This fourth order linear ordinary differential equation can be seen to factorize and so we find solutions are given by solutions of the two second order ordinary differential equations
\begin{equation}
\mathcal{L}[\phi_j] = 0,
\quad\textnormal{and}\quad
 \frac{1}{\Rey}\frac{\mu_j^*}{\rho_j^*}\mathcal{L}[\phi_j] = \textrm{i}\omega\phi_j.
\end{equation}
Hence, the general solution to \eqref{eq:osVisc} is given by
\begin{equation}
\phi_j = c_{j1}r^{-m} + c_{j2}r^m + c_{j3}\mathcal{J}_m\left(q_j\omega^{1/2}r\right) + c_{j4}\mathcal{Y}_m\left(q_j\omega^{1/2}r\right),
\end{equation}
where $j=1, 2$ corresponds to the inner and outer fluid layers respectively, $\mathcal{Y}$ is a Bessel function of the second kind and $q_j = (1-\textrm{i})\sqrt{\Rey\, \rho_j/2\mu_j}$.  The pressure perturbation is given by
\begin{equation} \label{eq:viscP}
P_j(r) = -\rho_j\left(2\phi_j - \frac{\omega r}{m}\phi_j'\right) +\frac{\mu_j}{\Rey} \frac{\textrm{i}r}{m}\left\{\phi_j''' + \frac{\phi_j''}{r} -\frac{(1+m^2)\phi_j'}{r^2} + \frac{2m^2\phi_j}{r^3}\right\}.
\end{equation}

Velocity regularity at $r=0$ forces $c_{11} = 0$ and $c_{14}=0$.  The no-slip and no-penetration conditions at $r=1$ can be enforced by taking
\begin{subequations}  \label{eq:noSlipnoPen}
\begin{gather}
c_{21} = -\frac{q_2\sqrt{\omega}}{2m}\left(\mathcal{J}_{m+1}(q_2\sqrt{\omega})c_{23} + \mathcal{Y}_{m+1}(q_2\sqrt{\omega})c_{24}\right)\\
c_{22} = -\left(\mathcal{J}_{m}(q_2\sqrt{\omega})c_{23} +\mathcal{Y}_{m}(q_2\sqrt{\omega})c_{24}\right)  + \frac{q_2\sqrt{\omega}}{2m}\left(\mathcal{J}_{m+1}(q_2\sqrt{\omega})c_{23} + \mathcal{Y}_{m+1}(q_2\sqrt{\omega})c_{24}\right).
\end{gather}
\end{subequations}
The kinematic condition at the interface is as for the inviscid case and so we require \eqref{eq:twoLayerKin} to hold.
We also require continuity of tangential fluid velocity across the interface for viscous fluids, and so we require \eqref{eq:tanCont} as well.

Substituting into the stress continuity conditions \eqref{eq:stressCont2} using \eqref{eq:viscP}, the interfacial jump conditions are, in the absence of surface tension, in the tangential and normal directions respectively
\begin{subequations} \label{eq:viscnost}
\begin{equation} \label{eq:viscnost1}
\left[\mu^*\left(r_0^2\phi'' - r_0\phi' + m^2\phi\right)\right]^+_- = 0,
\end{equation}
\begin{equation} \label{eq:viscnost2}
\left[\rho^*\left(r_0 - 2\phi + \frac{\omega r_0}{m}\phi'\right) + \frac{\mu^*}{\Rey}\frac{\textrm{i}r_0}{m}\left\{\phi''' - \frac{3m^2}{r_0^3}
\left(r_0\phi' - \phi\right)\right\}\right]^+_- = 0,
\end{equation}
\end{subequations}
where \eqref{eq:viscnost1} has been used to simplify \eqref{eq:viscnost2}.

In the special case that the dynamic viscosity of each layer is equal, then $\mu_j=1$ for $j=1,2$ and we can make use of the continuity of fluid velocity condition, that $\phi_j$ and $\phi_j'$ are continuous across the interface, to simplify \eqref{eq:viscnost} as
\begin{subequations} \label{eq:stressCont3}
\begin{equation} 
\left[\phi''\right]^+_- = 0,
\end{equation}
\begin{equation}
\left[\rho^*\left\{m + \omega \left(\phi'-2\right)\right\} + \frac{\textrm{i}}{\Rey}\phi'''
\right]^+_- = 0.
\end{equation}
\end{subequations}
To ensure continuity of velocity and stress and satisfy kinematic conditions at the interface in this special uniform viscosity case we therefore have the following three conditions:
\begin{subequations} \label{eq:viscInterfaceConditions}
\begin{gather}
\phi_1(r_0) = \frac{\omega r_0}{m} \\
\phi_2(r_0) = \frac{\omega r_0}{m} \\
\phi_1'(r_0) = \phi_2'(r_0), 
\end{gather}
\end{subequations}
and then two further conditions from either \eqref{eq:viscnost} if there is a viscosity contrast, or the simpler \eqref{eq:stressCont3} if the viscosities are uniform and equal.  The first four conditions can be used to find expressions for $c_{12}$, $c_{13}$, $c_{21}$, and $c_{22}$ in terms of  $\omega$, $\phi_1(r_0)$, $\phi_1'(r_0)$, $\phi_2(r_0)$ and $\phi_2'(r_0)$.  The final condition must be satisfied too and this yields the dispersion relation.

\begin{figure}
\begin{center}
\hspace*{-6pt}\includegraphics[clip = TRUE]{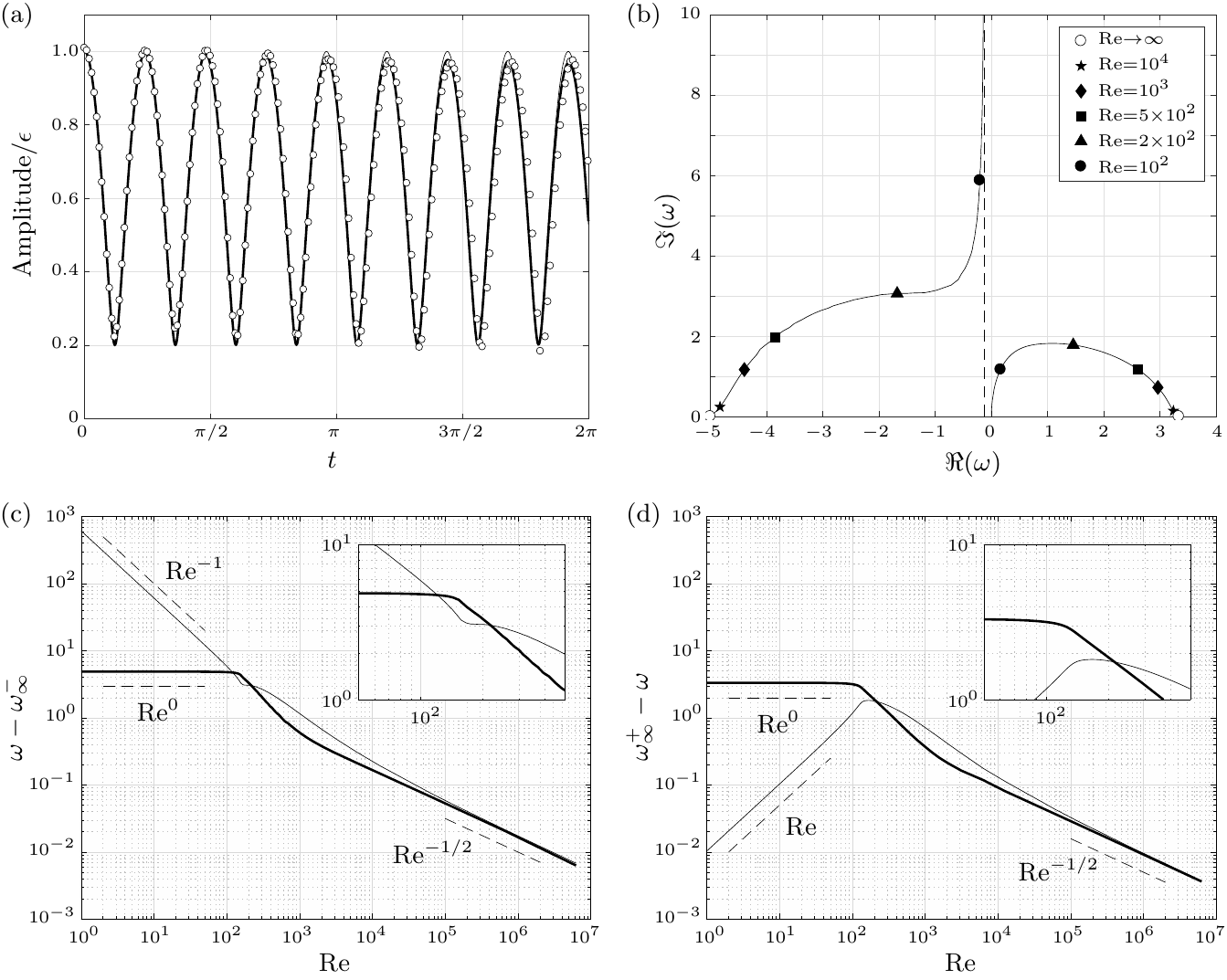}
\end{center}
\caption{\label{fig:rt4Figure10} The parameters for the simulations shown are $\mathscr{A} = \frac{5}{6}$, $m=20$, $r_0 = 2^{-1/2}$ and where viscous $\Rey = 2\pi\times10^6$. (a) The predicted normalised amplitude of a perturbation to the interface against time shown for the viscous analytical prediction (bold solid), the inviscid theoretical prediction (thin solid) and the numerical simulation (white data points).  The amplitude comprises the superposition of two waves, one travelling in the positive $\theta$ direction and the other in the negative $\theta$ direction. (b) The plot shows how the eigenvalues of the two modes of solution are modified as the viscosity of the two fluid layers is increased.  The inviscid solution ($\omega_\infty^\pm$, white data points) is given by \eqref{eq:disprel1}.  As the Reynolds number decreases the positive solution approaches zero, whereas the negative solution asymptotes to a finite real value (dashed vertical line) and infinite imaginary value.  The real value of the asymptote is given by \eqref{eq:imagAsy}. (c) The behaviour of the negative solution as the Reynolds number is varied.  In the small Reynolds number limit the solution behaves as $\omega \sim \lambda + \textrm{i}\kappa\Rey^{-1}$.  The real part of the solution is shown in bold and the imaginary part is shown as a thin line.  The power law behaviour of the solutions is indicated by the dashed lines.  (d) The behaviour of the positive solution as the Reynolds number is varied.  In the small Reynolds number limit the solution behaves as $\omega\sim \textrm{i}\kappa\Rey$ (see \eqref{eq:posRootAsy}).}  
\end{figure}

Figure \ref{fig:rt4Figure10} shows the behaviour of stable configurations of viscous fluid layers of uniform but differing density and a sharp interface.  The viscosity of the fluid layers is uniform and equal.  For the plots shown the Atwood number is $\mathscr{A} = \frac{5}{6}$, the azimuthal wavenumber is $m=20$, the initial interface position is $r_0 = 2^{-1/2}$, giving equal fluid volume in each layer.  Figure \ref{fig:rt4Figure10}a is a plot of the amplitude of the interfacial disturbance that propagates around the interface with time.  The white circles are taken from a low-viscosity numerical simulation with $\Rey = 2\pi\times10^6$.  The data points are well-modelled by the inviscid solution \eqref{eq:invSol} which comprises a superposition of two counter-propagating interfacial waves (thin solid line).  Closer agreement is found by using the full viscous solution that better captures the slow decay in amplitude of the perturbations with time (thick solid line).  Figure \ref{fig:rt4Figure10}b shows the behaviour of the eigenvalue $\omega$ as the viscosity of the fluid layers varies.  The inviscid solutions, $\omega_\infty^\pm$, given by \eqref{eq:invSol}, are shown as white circles.  The positive solution $\omega_\infty^+\approx 3.33$ corresponds to a wave that precesses in a clockwise direction about the interface whereas the negative solution $\omega_\infty^-\approx-5.00$ corresponds to a wave that precesses in an anticlockwise direction about the interface.  As the viscosity of the fluid layers is increased the precession rates of the solutions decreases, the waves are less able to propagate around the interface.  The positive solution tends to zero as $\Rey\to0$ representing a stationary perturbation.  The negative solution degenerates as its imaginary part tends to positive infinity, meaning the solution decays instantaneously.

We can examine the behaviour of the viscous solutions in the limit of small Reynolds number.  The root associated with $\omega_\infty^+$ behaves as
\begin{multline} \label{eq:posRootAsy}
\omega\sim\frac{\textrm{i}\mathscr{A}m}{2}
\left\{1-\frac{r_0^{2(m-1)}}{2}\left[\left(m^2(1-r_0^2)^2 + 2 r_0^2\right)(1 - \eta) + m(1 - r_0^4)(1 + \eta)\right] - \eta r_0^{4m}\right\}
\\
\times\frac{r_0^2}{m^2-1}
\left\{1 + \eta r_0^{2(m-1)}(m^2 (1-r_0^2)^2 + 2r_0^2) + \eta^2 r_0^{4m}\right\}^{-1}\Rey
+O(\Rey^2),
\end{multline}
where $\eta = (\mu_2 - \mu_1)/(\mu_2 + \mu_1)$ is the viscosity contrast \citep[cf.~(18)][noting that they have labelled their fluid layers the other way round to the present authors]{alvarez-lacalle}.  We can see that this mode is stable for positive Atwood number, $\mathscr{A}>0$, and unstable for negative Atwood number, $\mathscr{A}<0$.  When the Atwood number is negative, the magnitude of the growth rate is determined largely by the system Reynolds number, $\Rey$, the Atwood number, $\mathscr{A}$, and the azimuthal wavenumber, $m$.  The dependence on the viscosity contrast $\eta$ is multiplied by terms involving $r_0$ to powers of $2(m-1)$ or higher.  As $r_0\in(0,1)$ this indicates therefore, for moderate values of $m$, the dependence of the growth rate on $\eta$ may be weak and the growth rate for most modes of instability may depend largely only the viscosity of the most viscous layer, but it does not matter whether this layer is the inner layer or the outer layer. 

For the negative root associated with $\omega_\infty^-$, we can show $\omega \sim \lambda  + \textrm{i}\kappa\Rey^{-1}$ where the expressions for the coefficients $\kappa$ and $\lambda$ are unwieldy, but are given in appendix \ref{sec:lrnsb}.  Figures \ref{fig:rt4Figure10}c and \ref{fig:rt4Figure10}d show the asymptotic behaviours of the real and imaginary parts of the eigenvalue $\omega$ as the Reynolds number is varied.  Both solutions approach the inviscid solution like $\Rey^{-1/2}$ as $\Rey\to\infty$.  In the high viscosity limit as $\Rey\to0$ the positive solution tends to zero as shown in \eqref{eq:posRootAsy}, the negative solution tends to a constant real part and a singular imaginary part as described in appendix \ref{sec:lrnsb}.

In order for the system at low Reynolds number to support Saffman-Taylor instability \citep{saffmanTaylor} we would require there to exist an $\eta=\eta^*\in(-1, 1)$ such that $\omega=0$ in \eqref{eq:posRootAsy}, the neutral stability case, i.e., there would exist for $m\geqslant 2$ an $\eta^*$ that satisfies
\begin{equation}
\eta^* = \frac{2 - r_0^{2(m-1)}\left[m^2\left(1-r_0^2\right)^2+m(1-r_0^4)+2r_0^2\right]}{2r_0^{4m}-r_0^{2(m-1)}\left[m^2\left(1-r_0^2\right)^2-m(1-r_0^4)+2r_0^2\right]} \in[-1, 1],
\end{equation}
however no such $\eta^*$ exists, and so the system does not support Saffman-Taylor instability at low Reynolds number.

\begin{figure}
\begin{center}
\hspace*{-4pt}\includegraphics[clip = TRUE]{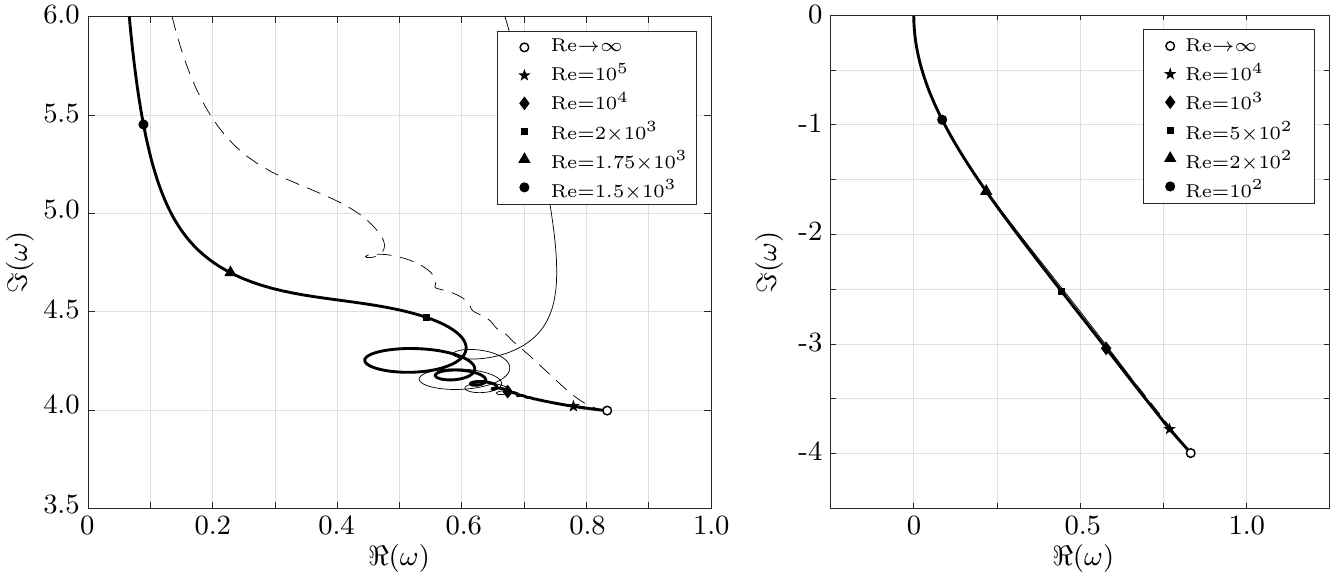}
\end{center}
\caption{\label{fig:unstabRTI}The plots show the behaviour of the eigenvalue $\omega$ with changes in the system Reynolds number, $\Rey$, for $\mathscr{A} = -5/6$, $m = 20$, $r_0 = 2^{-1/2}$.  The variation with the viscosity contrast is also shown: the bolid solid lines are $\eta = 0$, no contrast, matched viscosities; the thin solid lines are $\eta = 5/6$, the outer layer is 11 times more viscous than the inner layer; the dashed lines are $\eta = -5/6$, the inner layer is 11 times more viscous than the outer layer.  The stable root, left, has the same asymptotic behaviour as the negative root in figure \ref{fig:rt4Figure10}b, diverging to a positive infinite imaginary part as $\Rey^{-1}$ in the limit $\Rey\to0$. In the limit of large Reynolds number the solution tends toward the inviscid solution shown by the white circle.  The unstable solution tends to zero as $\Rey\to0$ and tends to the inviscid solution as $\Rey\to\infty$, shown by the white circle.  The asymptotic expression \eqref{eq:posRootAsy} agrees well with the numerical solution for near the $\Rey\to0$ limit.  While the stable branch is sensitive to the viscosity contrast, $\eta$, the unstable branch is insensitive for all $\Rey$; the differences between the three lines cannot be observed easily at the scale shown.}
\end{figure}

We now turn our attention to viscous fluid layers separated by an initially sharp interface with an unstable density stratification.  We consider the unstable complement of figure \ref{fig:rt4Figure10} where, as before, $m = 20$, $r_0 = 2^{-1/2}$, but we reverse the sign of the Atwood number such that the system is unstable and $\mathscr{A}=-5/6$.  Figure \ref{fig:unstabRTI} shows the behaviour of solutions to the system as the system Reynolds number is varied for three different values of the viscosity contrast, $\eta$.  The inviscid solutions \eqref{eq:invSol} are shown as white circles and it can be seen that as $\Rey\to\infty$ all the solutions tend toward their inviscid limit.  The numerical solutions tend toward the inviscid solution as $\Rey^{-1/2}$ as in figure \ref{fig:rt4Figure10}.  The stable, decaying, solutions are shown on the left and have the same behaviour asymptotically as $\Rey\to0$ as the negative solution in figure \ref{fig:rt4Figure10}, specifically that $\omega\sim\lambda + \textrm{i}\kappa \Rey^{-1}$.  The unstable solutions determine the growth of the perturbation at the interface and can be seen to tend to zero, the stationary solution, as $\Rey\to0$.  The solution near $\Rey\to0$ is well-approximated by \eqref{eq:posRootAsy}.  For the solutions shown the density contrasts were: $\eta = 0$, for the bold solid lines; $\eta = 5/6$ for the thin solid lines, and $\eta=-5/6$ for the dashed lines.  A contrast of $\eta = 5/6$ corresponds to the outer layer being 11 times more viscous than the inner layer, whereas a contrast of $\eta = -5/6$ corresponds to the inner layer being 11 times more viscous than the outer layer.  It can be seen that while the stable solution is sensitive to the viscosity contrast, the unstable solution, which controls the Rayleigh-Taylor growth of the perturbation, is insensitive to the value of $\eta$ across all values of $\Rey$.  This supports our earlier conjecture that the growth rate is determined primarily by the system Reynolds number and it does not matter whether the outer layer is more viscous than the inner layer or vice-versa.  In all cases we observe that as the system Reynolds number increases the growth rate of the perturbation is reduced.

\begin{figure} 
\begin{center}
\includegraphics[clip = TRUE]{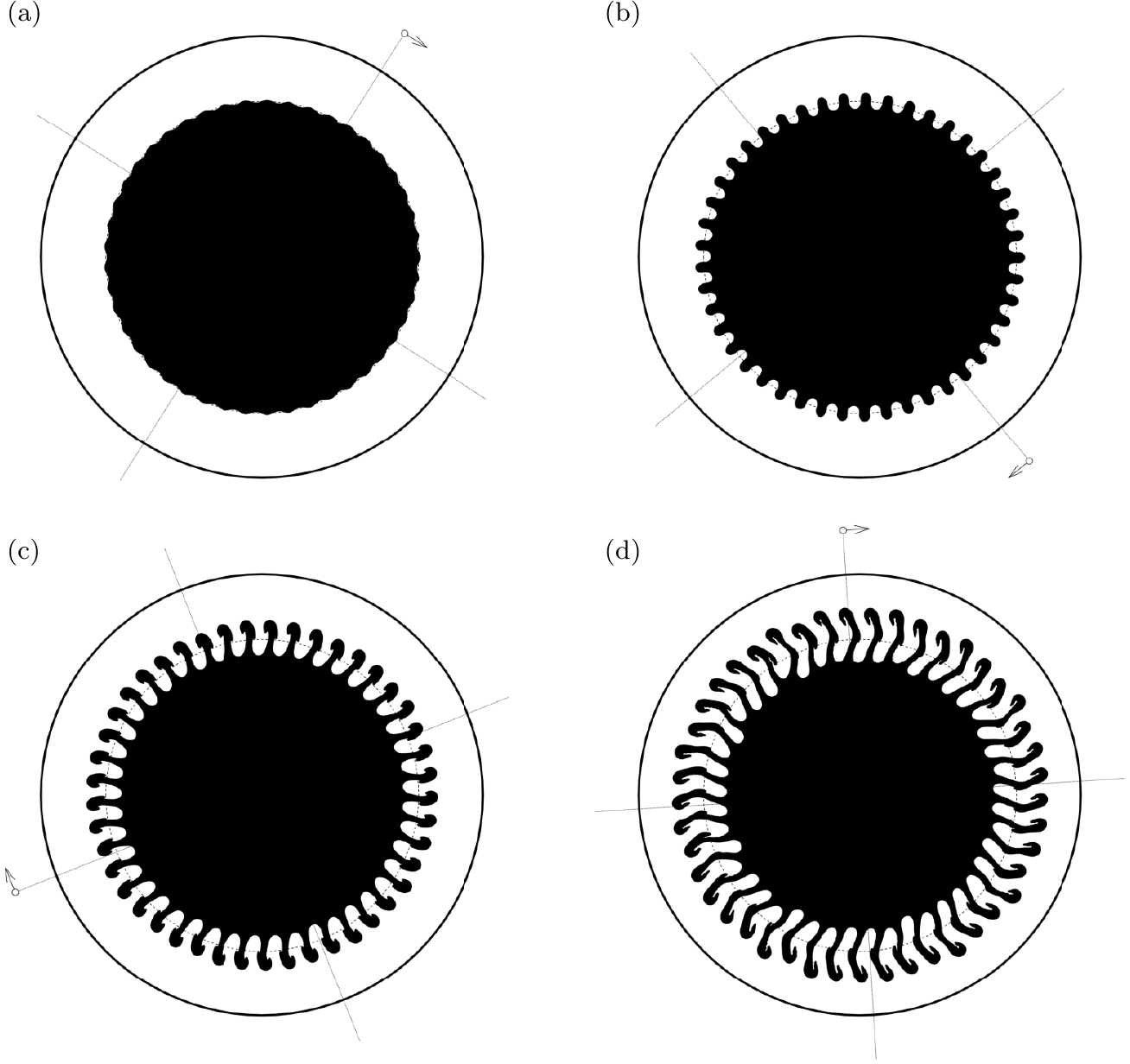}
\end{center}
\caption{\label{fig:rtiSim}Simulation of a centrifugally forced Rayleigh-Taylor instability.  The parameters were $\textrm{Re} = 2\pi\times10^2$, $\mathscr{A} = -\frac{1}{2}$, $r_0 = 2^{-1/2}$ (shown dashed), $\eta=0$, and the initial perturbation was $m=45$, $\epsilon = 4\times10^{-3}$.  The times shown are: (a) $t=0.57$, (b) $t=2.45$, (c) $t=4.34$, and (d) $t = 6.22$. }
\end{figure}

Figure \ref{fig:rtiSim} shows snapshots of a numerical simulation of two fluids of equal volume and equal viscosity, where $\mathscr{A} = -\frac{1}{2}$, $r_0 = 2^{-1/2}$, $\Rey = 2\pi\times10^2$, $\eta=0$ and the inital perturbation has azimuthal wavenumber $m=45$.  The initial amplitude of the perturbation was $\epsilon = 4\times10^{-3}$.  The four snapshots are from $t=0.57$ to $t = 6.22$, approximately one complete revolution of the system.  The boundary of the domain is shown as a thick solid line and the initial position of the interface is the inner dashed circle in each image.  The images are in the rotating frame of reference.  The large dotted cross with a white circle at one end indicates the position of a fixed point in the `laboratory', or fixed, inertial frame.  This fixed point appears to move in a clockwise direction.  As may be observed, the initial perturbation starts to grow.  It can be seen in the first two images that the growth is dominated by motion in the radial direction, the effect of the Coriolis term is seen to be small at these times, again the growth of the instability is driven by centrifugal forces.

\begin{figure} 
\begin{center}
\hspace*{-30pt}\includegraphics[clip = TRUE]{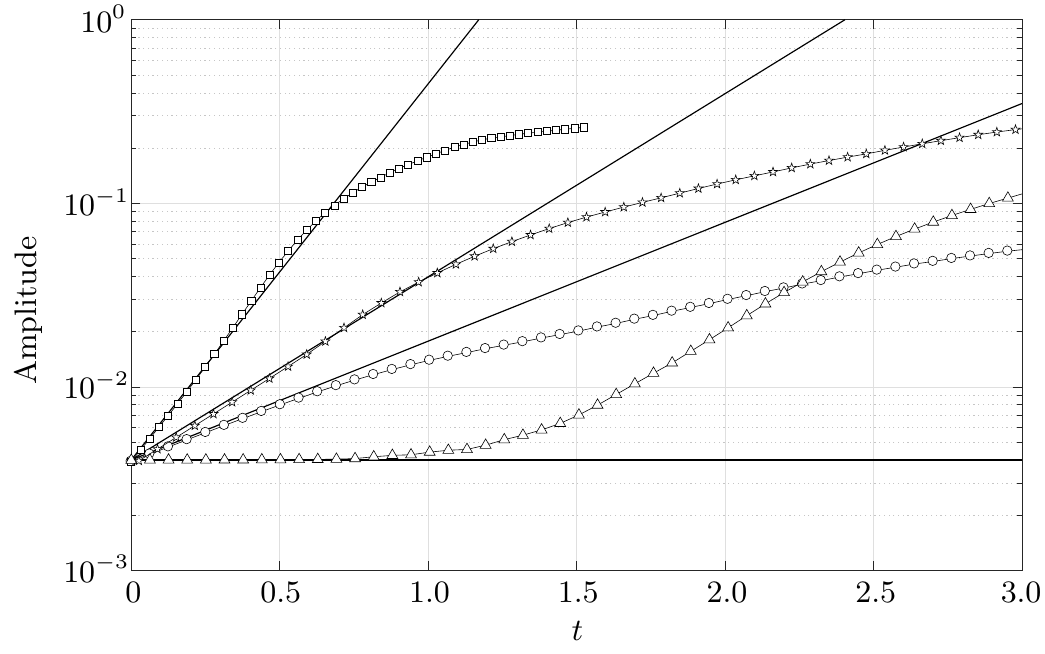}
\end{center}
\caption{\label{fig:theComp}A comparison of the growth rates observed in numerical simulations compared to the theoretical predictions.  The parameters for all simulations were $\mathscr{A} = -\frac{1}{2}$, $m = 45$, $r_0 = 2^{-1/2}$.  The individual simulation runs had: $({\scriptstyle\square})$ low viscosity $\Rey = 2\pi\times10^6$, no surface tension, no diffusion; $(\raisebox{-1pt}{\textbf{\scriptsize{\text{\FiveStarOpen}}}})$  low viscosity $\Rey = 2\pi\times10^6$, no surface tension, an initial diffuse interface with width $\delta = 0.067$; $(\circ)$ high viscosity $\Rey = 2\pi\times10^2$, no surface tension, no diffusion; (${\scriptstyle\triangle}$) low viscosity $\Rey = 2\pi\times10^6$, surface tension $\We = 5\times10^3$, no diffusion.  The straight thin lines are the theoretically predicted growth rates.}
\end{figure}

Figure \ref{fig:theComp} is a comparison of various growth rates calculated from numerical simulations of the centrifugally-driven Rayleigh-Taylor instability against the linear stability analysis predictions.  For parameters $\mathscr{A}=-\frac{1}{2}$, $r_0 = 2^{-1/2}$, $m=45$, $\eta=0$ and $\Rey= 2\pi\times10^6$, the growth rate is indicated by the square data points.  To a good approximation the data lies at early times on the straight line given by the inviscid approximation $\omega = 0.50 - 4.72\textrm{i}$.  For a simulation at more moderate Reynolds numbers we expect the inviscid approximation to be a poor estimate of the growth rate.  The white-circle data points are from a simulation with $\Rey=2\pi\times10^2$.  We may not neglect the effects of viscosity at this Reynolds number, as the growth rate is seen to be substantially lower than that of the high Reynolds number simulation.  The theoretical prediction for this configuration is that $\omega = 0.09-1.49\textrm{i}$, and can be seen to match well the numerical data points (white-circle).  The remaining two sets of data are for configurations with a diffuse interface and with surface tension which are discussed in \S\,\ref{sec:dvl}.

\subsection{\label{sec:dvl}Immiscible viscous layers with surface tension and miscible diffuse viscous fluid layers}


Two classes of viscous flow are of practical interest.  We may have a configuration where the two fluids are different, immiscible with different viscosities and surface tension between them.  We may also consider the case of miscible fluids that have approximately equal viscosities, no surface tension between the layers, but a diffuse interface.  

We first consider the case of two immiscible fluids with a constant density and constant viscosity contrast between them and surface tension acting at the interface.  The governing version of the Orr-Sommerfeld equation \eqref{eq:os} is as in \S\,\ref{sec:vfl} and is given by \eqref{eq:osVisc}, which accepts the same solution as before.  However, the boundary conditions at the interface must be modified to account for the stress jump due to the surface tension, similarly to the inviscid case considered in \S\,\ref{sec:surfaceTension}.  The kinematic condition (\ref{eq:viscInterfaceConditions}a, b), the tangential velocity continuity condition (\ref{eq:viscInterfaceConditions}c), and the tangential stress condition (\ref{eq:viscnost}a) remain as before.  However (\ref{eq:viscnost}b) is modified, following \eqref{eq:stressCont2} to give
\begin{equation} \label{eq:viscJump2}
\left[\rho^*\left(r_0 - 2\phi + \frac{\omega r_0}{m}\phi\right) + \frac{\mu^*}{\Rey}\frac{\textrm{i}{r}}{m}\left\{\phi''' - \frac{3m^2}{r_0^3}
\left(r_0\phi'-\phi\right)\right\}\right]^+_- = -\frac{1}{\We}\frac{m^2-1}{r_0^2},
\end{equation}
(cf.~the inviscid condition \eqref{eq:stJump}) where (\ref{eq:viscnost}a) has been used to simplify \eqref{eq:viscJump2}.  As in the miscible case, \eqref{eq:viscInterfaceConditions} and (\ref{eq:viscnost}a) may be used to find expressions for $c_{12}$, $c_{13}$, $c_{21}$, and $c_{22}$ in terms of $\omega$ and also then $\phi_1$ and $\phi_2$ and their gradients at $r_0$.  The final condition \eqref{eq:viscJump2} yields the dispersion relation.  The inclusion of the effect of surface tension in the right hand side of \eqref{eq:viscJump2} on the low Reynolds number asymptotic solution \eqref{eq:posRootAsy} is to introduce a correcting factor
$1+(m^2-1)/(2\We\,\mathscr{A}r_0^3) = 1 + S/m$ into the expression for $\omega$.  This has the corollary that the surface tension may stabilize the mode with azimuthal wavenumber $m^*$ when $S = -m^*$, or, equivalently, all modes $m>m^*$ are stabilized for $\We = -(m^{*2}-1)/(2\mathscr{A}r_0^3)$.

\begin{figure} 
\begin{center}
\hspace*{-20pt}\includegraphics[clip = TRUE]{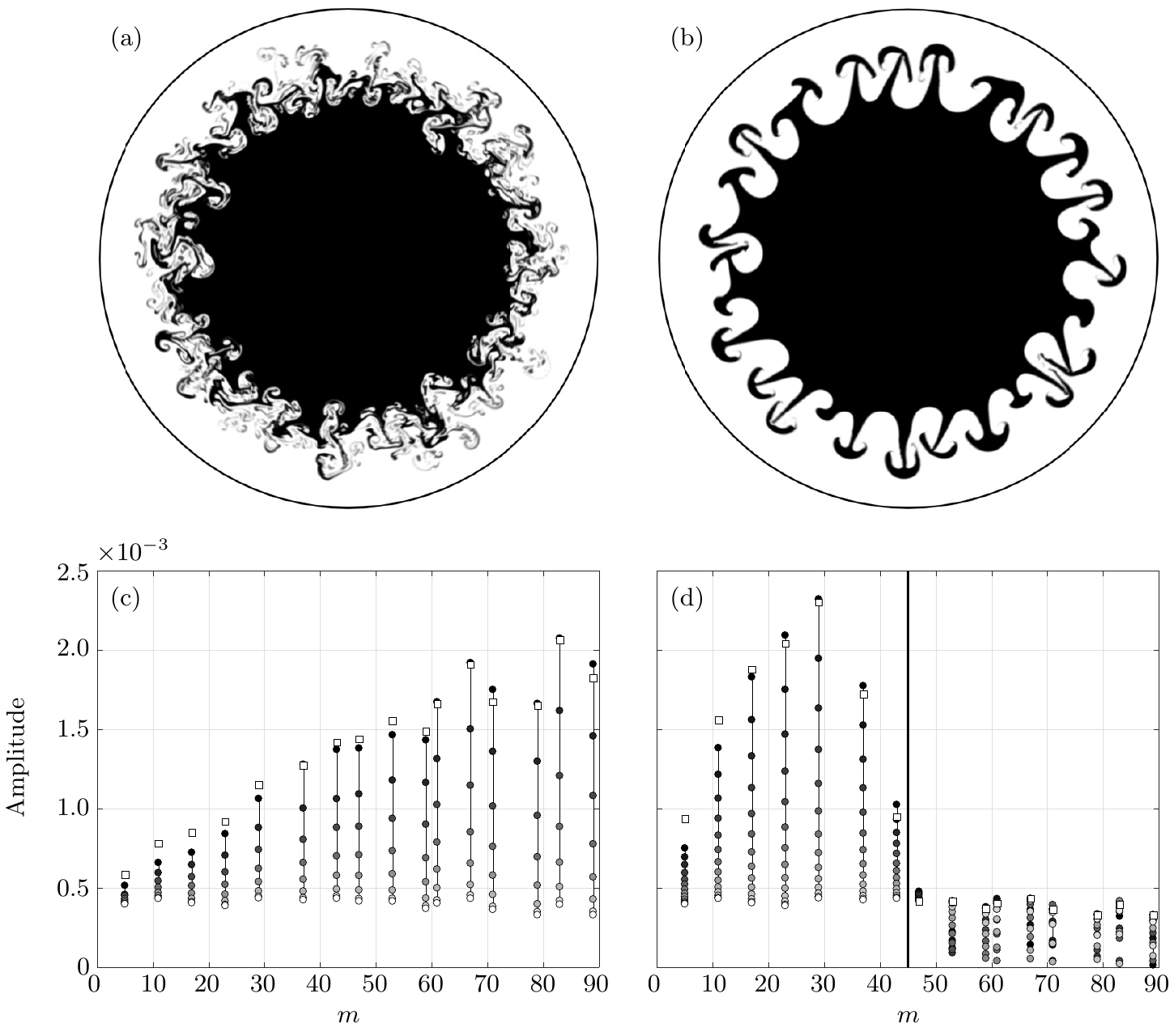}
\end{center}
\caption{\label{fig:stComp}The growth of modes of disturbance at an interface without surface tension (a, c) and with surface tension (b, d).  Snap shots from the flows are shown in images (a) and (b) at $t = 2.51$.  The surface tension in (b) and (d) correspond to a cut-off azimuthal wavenumber $m^*=45$ for growth as given by \eqref{eq:surfTenCutOff}, indicated by the bold vertical line.  In (c) and (d) it can be seen that each individual mode evolves in time, indicated by circular data points, filled white at $t=0$ and getting darker with each time step $\Delta t = 0.063$.  The predicted magnitude of each mode based on either \eqref{eq:disprel1} (c) or \eqref{eq:dispRelST2} (d), at the final time (black circle) is indicated by a white square.  The final times are $t=0.44$ in (c) and $t=0.82$ in (d).  It can be seen in (d) that modes that are above the cut-off wavenumber $m^*=45$ are suppressed at these early times, to a good approximation.  The parameters for the flow were $\mathscr{A} = -0.5$, and $\We = 5.8\times10^3$ in (b) and (d).}
\end{figure}

Figure \ref{fig:stComp} shows a comparison between the behaviour of two viscous systems, one without surface tension at the interface (a), and one with surface tension at the interface (b).  Both systems have equal fluid viscosities in each layer, i.e., $\eta=0$, and an unstable density stratification, $\mathscr{A}=-\frac{1}{2}$.  The Reynolds number in each simulation was $2\pi\times10^6$.  At $t=0$ a number of modes of perturbation between $m=5$ and $m=90$ were introduced at the interface with random amplitudes of order $4\times10^{-3}$.  This initial perturbation is indicated by the white circles on the graphs (c) and (d).  As time evolved and the instability developed, the amplitude of each mode was added to the graphs (c) and (d) as a progressively darker shaded data point for a given $m$ at time steps of $\Delta t = 0.063$.  At the last time shown ($t = 0.44$ in graph (c) and $t=0.82$ in graph (d)) the last data point (black circle) is compared with its theoretical prediction, indicated by a white square.  The simulation of the left, (a), had no surface tension acting at the interface, whereas the simulation on the right, (b), had surface tension acting at the interface and the effect can be immediately observed qualitatively.  The small-scale instabilities apparent in the left hand simulation appear significantly suppressed in the right hand simulation.  The strength of the surface tension, $\We=5800$, was chosen such that the critical wavenumber, $m^*$, given in \eqref{eq:surfTenCutOff} is approximately 45.  Image (d) shows that the behaviour of modes above $m=m^*$ is completely different to that below $m=m^*$ and their growth is suppressed at early times.

As the instability develops, some of the dense fluid moves towards the boundary of the domain, as is observed from the simulations.  As the interface between the two fluids moves towards the boundary it may develop areas where its radius of curvature is greater than its initial value at $r=r_0$.  This therefore allows modes that were suppressed initially by the surface tension when the interface was nearer the centre of the system to develop as the interface approaches the boundary since the effects of surface tension are not felt as strongly in regions of lower interfacial curvature.  We can return to figure \ref{fig:theComp} and consider the growth of a system where initially the perturbed mode is $m=45$, but the surface tension $\We=5\times10^3$ should completely suppress the growth of the mode.  The triangular data points show that initially the $m=45$ mode is unable to grow, but later as the interface has moved toward the boundary, due to instability at lower wavenumbers, the $m=45$ mode is able to grow.

We now turn to the case of two layers of miscible viscous fluid with matching viscosities whose interface has diffused over a period of time.  The approach is identical to that followed in the inviscid case where we consider a diffuse layer thickness $\delta\sim2\sqrt{\pi t/\textrm{Pe}}$ and a piecewise continuous density of the form \eqref{eq:densityStrat}.  Assuming equal viscosities for the fluids we define the following quantity
\begin{equation}
\zeta(r) = \frac{\Rey\, \beta\omega\textrm{i}}{\left(2+\alpha\right)^2}r^{2+\alpha},
\end{equation}
and functions
\begin{eqnarray}
\mathcal{F}(c;\zeta) &=& _2F_3\left(\left[\frac{\alpha+2(c+\chi)}{2\left(2+\alpha\right)}, \frac{\alpha+2(c-\chi)}{2\left(2+\alpha\right)}\right];
\left[\frac{\alpha}{2+\alpha},\frac{\alpha+2c}{2+\alpha},\frac{\alpha+2(c+1)}{2+\alpha}\right];\zeta\right) \qquad \\
\mathcal{G}(c;\zeta) &=& _2F_3\left(\left[\frac{\alpha+2(c+\chi)}{2\left(2+\alpha\right)}, \frac{\alpha+2(c-\chi)}{2\left(2+\alpha\right)}\right];
\left[\frac{4+\alpha}{2+\alpha},\frac{\alpha+2(c-1)}{2+\alpha},\frac{\alpha+2c}{2+\alpha}\right];\zeta\right)
\end{eqnarray}
where $_2F_3$ is a hypergeometric function and $\chi$ is as defined in \eqref{eq:gammadef}.  We may write solutions for each region as
\begin{equation}
\phi_1 = c_{11}r^{-m} + c_{12}r^m + c_{13}\mathcal{J}_m(q_1\omega^{1/2}r) + c_{14}\mathcal{Y}_m(q_1\omega^{1/2}r),
\end{equation}
\begin{equation}
\phi_2 = c_{21}r^{-m}\mathcal{F}(-m;\zeta) + c_{22}r^m\mathcal{F}(m;\zeta) + c_{23}r^{2-m}\mathcal{G}(2-m,\zeta) + c_{24}r^{2+m}\mathcal{G}(2+m,\zeta),
\end{equation}
\begin{equation}
\phi_3 = c_{31}r^{-m} + c_{32}r^m + c_{33}\mathcal{J}_m(q_2\omega^{1/2}r) + c_{34}\mathcal{Y}_m(q_2\omega^{1/2}r).
\end{equation}
We then proceed as before, enforcing velocity regularity at $r=0$ which forces $c_{11}=0$ and $c_{14}=0$, and no-slip, no-penetration conditions at $r=1$ which forces equivalent constraints to \eqref{eq:noSlipnoPen}.  This reduces the number of free constants to eight; two in the inner and outer layers, and four in the mid-layer.  We now apply the kinematic condition and velocity continuity conditions at each interface (five conditions) and stress continuity at each interface (four conditions)  to find the dispersion relation.

\begin{figure}
\begin{center}
\hspace*{-20pt} \includegraphics[clip = TRUE]{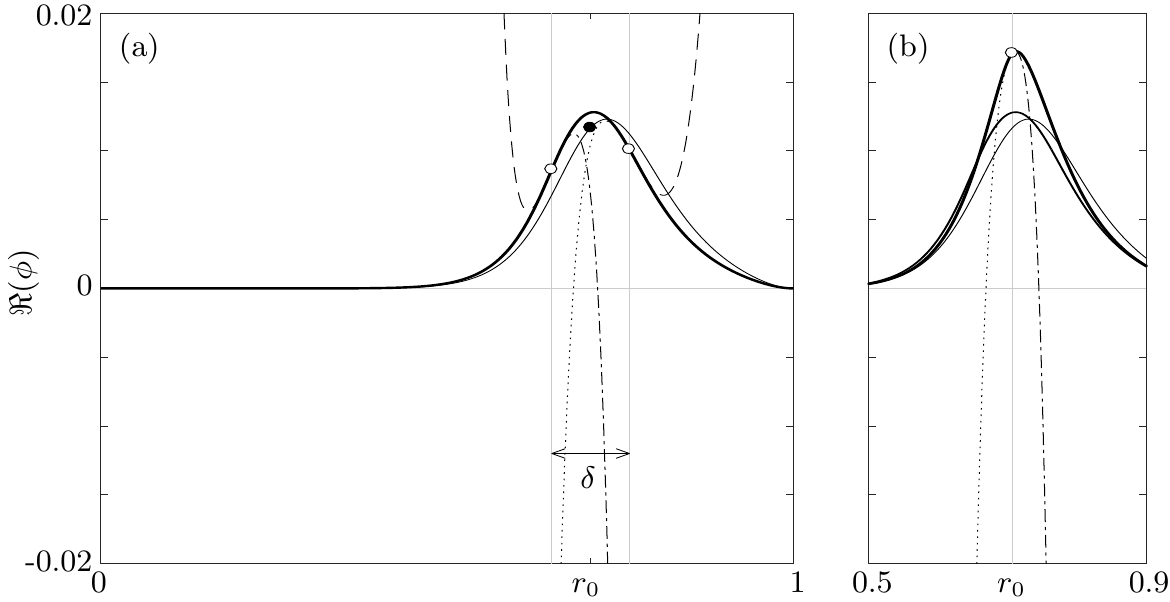}
\end{center}
\caption{\label{fig:viscDiff}(a) The solution, $\phi$, (solid line) to \eqref{eq:os} for $m=12$, 
$\mathscr{A} = -\frac{1}{2}$, $r_0 = 2^{-1/2}$, $\textrm{Re} = 400$, $\mu_1=1$, $\mu_2=1$, and $\delta = 2\sqrt{\pi t/\Pe}\approx0.11$, where $t = 10^{-3}\Pe$.  The shown solution is the real part, and corresponds to the solution of the dispersion relation at $\omega = 0.22 - 1.41\textrm{i}$.  The solution constitutes the matching of three analytic solutions; $\phi_1$ for $r\in[0,r_{0-}]$ which continues dot-dashed after the solution is matched to $\phi_2$ for $r\in[r_{0-}, r_{0+}]$.  Outside this region $\phi_2$ is shown dashed.  Finally in $r\in[r_{0+},1]$, $\phi_3$ is used, but is shown dashed outside this range.  The diffusion width $\delta$ is indicated, and the matching between solutions occurs at the white dots. (b) Comparison with the zero diffusion case (higher amplitude curve); again the transition between the two solutions is shown with a white dot, the solution corresponds to the solution of the dispersion relation at $\omega = 0.29 - 1.62\textrm{i}$.}
\end{figure}

Figure \ref{fig:viscDiff} shows the effect of a diffuse interface on the real part of the solution $\phi$.  The piecewise solution comprises three segments that, as in the inviscid case, transition at the white circles.  The unused parts of the solutions are shown, indicating their behaviour in the matching regions.  The exact solution is shown as a thin solid line, the kinematic condition is enforced at $r = r_0$, indicated by the black circle.  The solution to \eqref{eq:os} subject to $\phi(0) = \phi'(0) = 0$, $\phi(1) = \phi'(1) = 0$ was calculated separately in $[0, r_0]$ and $[r_0, 1]$ with the kinematic condition enforced for each solution at $r=r_0$.  The free boundary condition $\phi'(r_0)$ in each domain and the eigenvalue $\omega$ are chosen to ensure that $\phi$ is at least class $C^4[0, 1]$.  The method of solution, as in the inviscid case, was to formulate the problem in terms of Chebyshev polynomials.  In plot (b) the comparison with the non-diffuse case is shown and it can be seen that the qualitative differences between the solutions are less marked than in the inviscid case since the viscosity has acted to smooth $\phi$ to some extent already.

\subsection{Non-inertial flow and comparison with Hele-Shaw cell and porous media flow}


A related problem to those considered above is of a two-dimensional rotating droplet, possibly lying within an unbounded fluid of different density and viscosity, rotating in a Hele-Shaw cell or a porous medium.  This problem was considered by \citet{schwartz} and later by \citet{alvarez-lacalle}.  If the outer fluid is considered unbounded and inertia is to be ignored then the nondimensionalisation used above is not natural since $a\to\infty$, but $\rho_0\Omega a^2/\mu_0\to 0$.  It is more appropriate therefore to take the radius of the `droplet', $r_0$, as the length scale in this case.  

We make a Stokes flow approximation, taking $\Rey\to0$ whereby inertial terms are ignored but the centrifugal term and the pressure term are balanced with the viscous term of the equation of motion \eqref{eq:Mom1}.  This requires a velocity scale $\left(\rho_0\Omega r_0^2/\mu_0\right)\Omega r_0$ and hence an implied time scale $\mu_0/\left(\rho_0\Omega^2r_0^2\right)$.  We also require
the pressure scale $\rho_0 \Omega^2 r_0^2$, and define a Reynolds number $\Rey = \left(\rho_0\Omega r_0^2/\mu_0\right)^2$.  The nondimensional equations of motion are, assuming constant viscosity and density in each layer
\begin{subequations} \label{eq:nond2}
\begin{equation}
\pd{\rho_j'}{t'} + \nabla'\cdot\left(\rho'_j\vc{u}'_j\right) = 0,
\end{equation}
\begin{equation}
\Rey\Dv{\vc{u}'_j}{t'} = -\frac{1}{\rho'_j}\nabla' p'_j +r'\hat{\vc{r}} - 2\Rey^{1/2}\hat{\vc{z}}\times\vc{u}'_j + \frac{\mu'_j}{\rho'_j}\nabla'^2\vc{u}'_j,
\end{equation}
\begin{equation}
\nabla'\cdot\vc{u}'_j=0.
\end{equation}
\end{subequations}
Dropping the prime notation and taking the Stokes flow limit $\Rey\to0$ we have that in each layer
\begin{subequations} \label{eq:nond3}
\begin{equation}
\pd{\rho_j}{t} + \nabla\cdot\left(\rho_j\vc{u}_j\right) = 0,
\end{equation}
\begin{equation}
\vc{0} = -\frac{1}{\rho_j}\nabla p_j +r\hat{\vc{r}} + \frac{\mu_j}{\rho_j}\nabla^2\vc{u}_j,
\end{equation}
\begin{equation}
\nabla\cdot\vc{u}_j=0.
\end{equation}
\end{subequations}
Under this approximation the convective derivative of the velocity and the Coriolis term are negligible.  We substitute for the fluid density, velocity and pressure as in \eqref{eq:SubsStart}--\eqref{eq:ueq} and find the corresponding zero-inertia Orr-Sommerfeld equation is
\begin{equation} \label{eq:noInertiaOS}
0 = \phi_j'''' + \frac{2\phi_j'''}{r} - \left(1+2m^2\right)\left[\frac{\phi_j''}{r^2} - \frac{\phi_j'}{r^3}\right] + \frac{m^2\left(m^2-4\right)\phi_j}{r^4} = \mathcal{L}^2[\phi].
\end{equation}
This is a simplification of \eqref{eq:os} in which the left hand side, the inertial terms are zero and we have uniform viscosity in each layer.  The corresponding pressure perturbation is given by
\begin{equation}
P_j = \frac{\textrm{i}\mu_j}{m}\left(\left(r\phi_j''\right)' - \frac{m^2+1}{r}\phi_j' + \frac{2m^2\phi_j}{r^2}\right).
\end{equation}
This zero-inertia form of the Orr-Sommerfeld equation accepts power-law solutions of the form
\begin{equation}
\phi_j = c_{j1}r^{-m} + c_{j2}r^{-m+2} + c_{j3}r^m + c_{j4}r^{m+2}, \quad j=1,2.
\end{equation}
The coefficients $c_{11}$ and $c_{12}$ are taken to be zero for velocity regularity at the origin.  The no-slip and no-penetration conditions are now applied at $r=a$ such that $\phi_2(a)=0$, $\phi_2'(a) = 0$.  The kinematic condition, which also enforces normal velocity continuity, is given now by $\phi_1(1) = \phi_2(1) = \omega/m$.  Tangential velocity continuity is ensured by setting $\phi_1'(1) = \phi_2'(1)$.  This leaves only one free constant, but the two stress continuity conditions to satisfy and so will yield a dispersion relation.  The normal and tangential stress continuity conditions are, at order $\epsilon$, respectively
\begin{gather}
\left[\mu^*\left\{\phi''' - 3m^2\left(\phi'-\phi\right)\right\} - \textrm{i}m\rho^* \right]^+_- =\frac{\textrm{i}m\left(m^2-1\right)}{\We},
\\
\left[\mu^*\left\{\phi'' - \phi' + m^2\phi\right\}\right]^+_- = 0,
\end{gather}
where the jump occurs across $r=1$, and the tangential condition has been used to simplify the normal condition.  For $a\gg1$, $m>1$, the dispersion relation is given by
\begin{multline} \label{eq:noInertiadr}
\omega = \frac{\textrm{i}\mathscr{A}m}{2}\left(1+\frac{1}{\We}\frac{m^2-1}{2\mathscr{A}}\right)
\\
\times\left(1-\frac{a^{-2(m+1)}}{2}\left[(a^2-1)^2(1-\eta)m^2+(a^4-1)(1+\eta)m+2a^2(1-\eta)\right]-a^{-4m}\eta\right)
\\
\times\left\{(m^2-1)\left[1+\eta\left(a^{-2(m+1)}\left[(a^2-1)^2m^2 + 2a^2\right]+a^{-4m}\eta\right)\right]\right\}^{-1},
\end{multline}
(cf.~\eqref{eq:posRootAsy} with the surface tension correction factor).  The form of the dispersion relation shows immediately that solutions will either decay or grow, depending on whether the Atwood number is positive or negative respectively, but there are no precessional or travelling wave solutions possible as might be anticipated on physical grounds for a system with no inertia.  There can be a balance between the stabilizing effect of the surface tension and the Rayleigh-Taylor instability such that a perturbed interface neither grows nor decays and is stationary.  This occurs when
\begin{equation}
\We = -\frac{m^2-1}{2\mathscr{A}},
\end{equation}
which can only occur for $\mathscr{A}<0$ as would be expected again on physical grounds.  If quantites are rewritten in terms of our original nondimensionalization (\S\,\ref{sec:eqMotion}) and $\mu_1 = \mu_2$, i.e., $\eta=0$, then the dispersion relation \eqref{eq:noInertiadr} is exactly the first term of \eqref{eq:posRootAsy} with the surface tension correction factor, as it must be.

We compare the results of \citet{schwartz} and \citet{alvarez-lacalle} with the related non-inertial limit of the flows considered in \S\S\,\ref{sec:crti}--\ref{sec:dvl}.  In a Hele-Shaw cell the equations of motion are classically simplified under Stokes flow and lubrication approximations that respectively ignore any inertia in the flow and assume that gradients in the gap-width direction are much greater than gradients in the planar direction.  For consideration of the flow in a Hele-Shaw cell or a porous media we return to \eqref{eq:nond2} but instead of interpretting the equations in plane polar coordinates we interpret them in cylindrical polar coordinates and apply a lubrication approximation whereby changes in the planar direction take place over much greater distances than changes in the gap-width $z$-direction.  That is, we apply a second scaling to \eqref{eq:nond2} under the assumption that the dimensional gap-width, $b$, is small compared to the initial radius of the inner fluid layer.  Hence, we take $\varepsilon = b/r_0 \ll1$.  Under this rescaling we write the gradient operator and the velocity separately in their planar and vertical components: $\nabla = \nabla_H+\partial\cdot/\partial z\hat{\vc{z}}$, $\vc{u} = \vc{u}_H + w\hat{\vc{z}}$.  The pressure, $p$, is scaled by $\varepsilon^{-2}$, and the system \eqref{eq:nond2} is transformed to
\begin{subequations}
\begin{equation}
\pd{\rho_j'}{t'} + \nabla'\cdot\left(\rho_j'\vc{u}_j'\right) = 0,
\end{equation}
\begin{equation}
\Rey\,\varepsilon^2\Dv{\vc{u}_{Hj}'}{t'} = -\frac{1}{\rho_j'}\nabla'_H p' + r'\hat{\vc{r}} - 2\Rey^{1/2}\varepsilon\,
\hat{\vc{z}}\times\vc{u}_{Hj}' + \frac{\mu_j'}{\rho_j'}\left(\varepsilon^2\nabla_H'^2+\partial_z^2\right)\vc{u}_{Hj}',
\end{equation}
\begin{equation}
\Rey\,\varepsilon^2\Dv{w_j'}{t'} = -\frac{1}{\varepsilon^2}\frac{1}{\rho_j'}\pd{p'}{z'} + \frac{\mu_j'}{\rho_j'}\left(
\varepsilon^2\nabla_H'^2 + \partial_z^2\right)w_j',
\end{equation}
\begin{equation}
\nabla'\cdot\vc{u}_j'=0.
\end{equation}
\end{subequations}
Hence, at leading order \eqref{eq:nond3} is replaced by
\begin{subequations}
\begin{equation}
\pd{\rho_j}{t} + \nabla\cdot\left(\rho_j\vc{u}_j\right) = 0,
\end{equation}
\begin{equation}
\vc{0} = -\frac{1}{\rho_j}\nabla_H p + r\hat{\vc{r}}+ \frac{\mu_j}{\rho_j}\pd{^2\vc{u}_{Hj}}{z^2},
\end{equation}
\begin{equation}
0 =\pd{p}{z},
\end{equation}
\begin{equation}
\nabla\cdot\vc{u}_j=0.
\end{equation}
\end{subequations}
It follows in the usual manner that $p = p(r, \theta, t)$ is independent of $z$ and so if the plates are located at $z=0$ and $z=\varepsilon$, the velocity field at leading order is
\begin{equation}
\vc{u}_j = \frac{z(z-\varepsilon)}{2\mu_j}\nabla_H\left(p_j - \frac{\rho_jr^2}{2}\right),
\end{equation}
and we can define a vertically averaged velocity in each fluid layer
\begin{equation} \label{eq:darcysLaw}
\vc{v}_j = \frac{1}{\varepsilon}\int_0^\varepsilon \vc{u}_j\,\textrm{d}z = -\frac{\varepsilon^2}{12\mu_j}\nabla_H\left(p_j - \frac{\rho_jr^2}{2}\right),
\end{equation}
a rotational version of Darcy's law.  As the pressure in each layer is independent of $z$ it follows from taking the curl of \eqref{eq:darcysLaw} that the vertically-averaged velocity, $\vc{v}_j$, is irrotational in each layer and so \citet{alvarez-lacalle} proceed by taking a potential for the flow such that $\vc{v}_j = \nabla\varphi_j$.  Hence, as a result of incompressibility, they seek to solve $\nabla^2_H\varphi_j=0$ in each layer.  For consistency with notation in the previous sections we follow an equivalent method which is to express $\vc{v}_j$ in terms of a streamfunction, $\psi_j$, since the velocity field is incompressible.  The irrotationality of $\vc{v}_j$ then implies that $\nabla_H^2\psi_j=0$.  We therefore seek normal mode solutions and solve Laplace's equation for $\psi_j=\epsilon \phi_j(r)\exp\{\textrm{i}(m\theta+\omega t)\}$ (taking care to distinguish between $\varepsilon$ the nondimensional gap-width of the Hele-Shaw cell, and $\epsilon$ the small perturbation to the background hydrostatic initial conditions). The pressure is then given by
\begin{equation}
p_j = p^*_j - \epsilon \frac{12\mu_j}{\varepsilon^2}\frac{\textrm{i}r}{m}\phi_j'\,\textrm{e}^{\textrm{i}(m\theta + \omega t)},
\quad\textnormal{where}\quad
p^*_j = p_{0j} + \frac{\rho_j r^2}{2},
\end{equation}
and $p_{0j}$ is a reference pressure.  (Note that solving Laplace's equation for $\psi$ leads to exactly the same equation for $\phi$ as the Rayleigh equation \eqref{eq:rayleighapprox} in \ref{sec:crti} demonstrating the well-known phenomena of three-dimensional viscous flow in a Hele-Shaw cell modelling two-dimensional inviscid flow.)  The conditions of stress continuity are
\begin{subequations}
\begin{eqnarray}
\left[p^*\right]^+_- = -\frac{1}{\We} \hspace*{34pt} && \textnormal{$O(1)$ normal stress}\\
\left[-\rho^* + \frac{2\textrm{i}\mu^*}{m}\left\{\left(\frac{6}{\varepsilon^2}+m^2\right)\phi' -m^2\phi\right\}\right]^+_- = \frac{1}{\We}\left(m^2 - 1\right) && \textnormal{$O(\epsilon)$ normal stress} \\
\left[\mu^*\left(\phi'' - \phi' + m^2\phi\right) \right]^+_- = 0, \hspace*{50pt} && \textnormal{$O(\epsilon)$ tangential stress} \hspace*{32pt}
\end{eqnarray}
\end{subequations}
where the jump occurs at $r=1$.  The kinematic condition is that $\phi_j(1)=\omega/m$, which also guarantees the continuity of the normal velocity across the interface.  To enforce continuity of tangential velocity we would require $\phi_1'(1) = \phi_2'(1)$.

The solutions to Laplace's equation have only two free constants in each layer, and once the conditions of regularity at the origin and unboundedness in the outer layer are taken into account, only two free constants remain to satisfy the tangential velocity continuity condition, stress continuity conditions and kinematic condition.  \citet{alvarez-lacalle} enforce the kinematic condition, and thus normal velocity continuity, and normal stress continuity to find (in our notation)
\begin{equation}
\phi_1 = \frac{\omega r^m}{m},
\quad
\phi_2 = \frac{\omega r^{-m}}{m},
\end{equation}
and hence find the dispersion relation
\begin{equation} \label{eq:aldr}
\omega = \frac{\textrm{i}\mathscr{A}m}{2}\left(1 + \frac{1}{\We}\frac{m^2-1}{2\mathscr{A}}\right)\left[\frac{6}{\varepsilon^2} + m\left(m + \eta\right)\right]^{-1}.
\end{equation}
We note that as in the plane-polar two-dimensional case, there can be a balance between the unstable density stratification and surface tension that leads to a stationary perturbation when $\We = -(m^2-1)/2\mathscr{A}$.  The conditions of tangential stress and velocity continuity, are not enforced.

\begin{figure}
\begin{center}
\includegraphics[clip = TRUE]{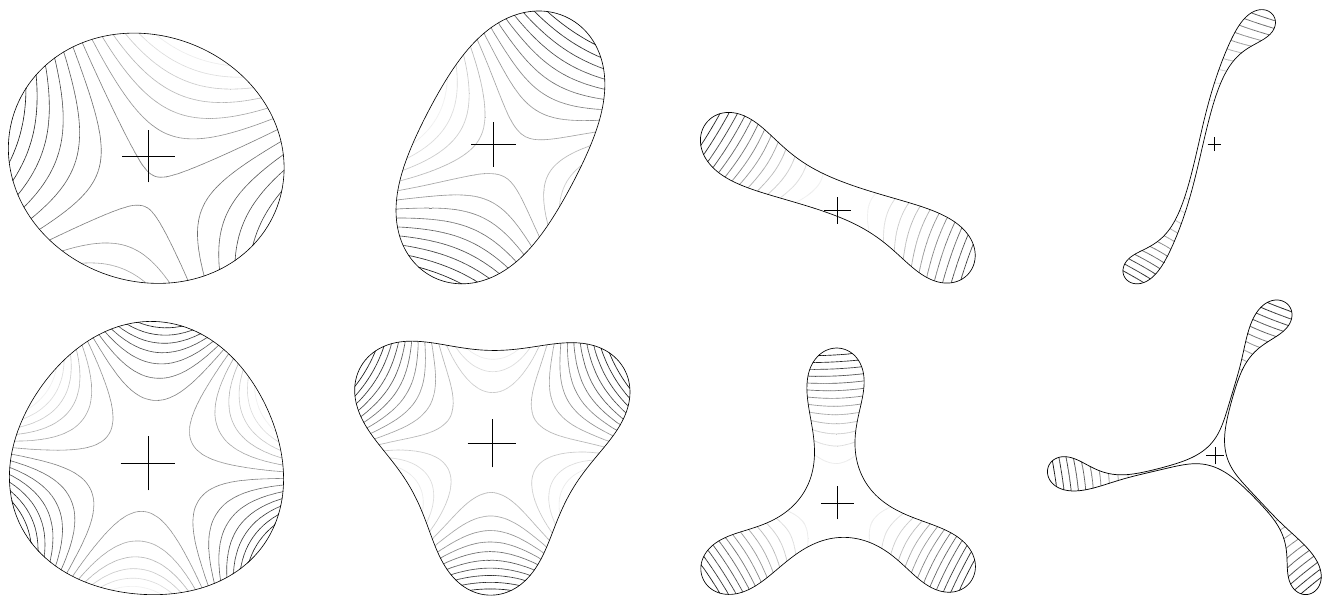}
\end{center}
\caption{\label{fig:schwartz}Modified pressure contours in perturbed droplets rotating in porous media or Hele-Shaw cells.  The initial droplet size is a perturbed circle of initially unit radius, and the cross is for scale (length 0.4 in each direction).  The times shown are $t = 15$, $25$, $30$, and $35$.  There is no external fluid layer.}
\end{figure}

\citet{schwartz} considered the same problem in a porous media as well as in a Hele-Shaw cell.  Figure \ref{fig:schwartz} shows droplets with differing initial perturbations in porous media evolving in time, governed by Schwartz's model; contours of $p-\rho r^2/2$ are shown.  Flows in Hele-Shaw cells of gap-width $b$ are equivalent to flows in porous media, governed by Darcy's Law, with permeability $k = b^2/12$.  Therefore for a fluid droplet, with no outer fluid layer, in an unbounded porous media we have $\mathscr{A}=-1$, $\eta = -1$ and it follows from \eqref{eq:aldr} that
\begin{equation} \label{eq:myschwartzdr}
\omega = -\frac{\textrm{i}m}{2}\left(1 - \frac{1}{\We}\frac{m^2-1}{2}\right)\left[\frac{1}{2k'}+m(m-1)\right]^{-1},
\end{equation}
where $k' = k/r_0^2$ is the nondimensional permeability.  The differences between \eqref{eq:myschwartzdr} (that agrees with the result of \citet{alvarez-lacalle}) and the result quoted in \citet{schwartz} are due to both applying normal stress continuity here, instead of pressure continuity, across the interface and not having a Coriolis term here for consistency with the narrow gap-width and Stokes flow approximations.  The dispersion relation given in \citet{schwartz} is equal to that of \citet{alvarez-lacalle} when $\Rey\to0$ in the Schwartz solution and pressure continuity is enforced in the \citet{alvarez-lacalle} solution (see their (13) as opposed to their (18)).  The behaviour of the plane-polar two-dimensional system and the Hele-Shaw cell, porous media solutions can be seen to have some qualitatively similar behaviour.  Comparing \eqref{eq:aldr} with \eqref{eq:noInertiadr} in the limit $a\to\infty$ which gives
\begin{equation}
\omega = \frac{\textrm{i}\mathscr{A}m}{2}\left(1+\frac{1}{\We}\frac{m^2-1}{2\mathscr{A}}\right)
\frac{1}{m^2-1},
\end{equation}
we see that the growth rates differ by a shape factor that depends on the azimuthal wavenumber and the viscosity contrast, but that the dependence on the Atwood number and the Weber number is identical.  The key difference between the models is the interaction with the viscous term in the equation of motion and so it is consistent that the behaviour of the models is distinguished via a shape factor that depends on viscosity and on the size of a given perturbation.

\section{\label{sec:conc}Conclusions}


We have considered perturbations to a two-dimensional system of concentric fluid layers in a circular domain that is undergoing constant rotation about its centre.  The fluids may differ in both density and viscosity and they may have a sharp interface possibly with surface tension acting, or they may have a diffuse interface.  We have carried out a linear stability analysis of the hydrostatic initial condition of the two fluids at rest in the rotating frame of reference.  When the inner fluid is less dense than the outer fluid the linear stability analysis shows the system to be stable, though this necessarily does not formally exclude the possibility of sub-critical instabilities existing.  When the inner fluid is denser than the outer fluid the system may be unstable.  When the system is unstable we see Rayleigh-Taylor-like growth driven by the rotation of the system.

As the fluids are assumed incompressible we posed descriptions of the velocities in the form of streamfunctions and sought a normal-mode decomposition for small perturbations to the hydrostatic background flow that yielded a fourth-order ordinary differential equation of an Orr-Sommerfeld type for each fluid layer.  We considered solutions in a variety of configurations.

The most straightforward configuration considered was for two inviscid fluids of differing, but uniform, density.  The Orr-Sommerfeld equation simplifies to a quadratic Rayleigh equation that accepts two power-law solutions.  The dispersion relation for this system was shown to be quadratic in $\omega$ and stable when $\mathscr{A}>0$, but unstable when $\mathscr{A}<0$.  The dispersion relation, via consideration of its discriminant, showed that any Rayleigh-Taylor growth was due to the centrifugal term in the equation of motion and hence the instability is `centrifugally driven'.  The effect of the Coriolis term when the system is unstable is always to inhibit the growth rate.  If the two fluids are immiscible and have some surface tension between them at the interface, the effect of the surface tension can be thought of formally in terms of modifying the apparent azimuthal wavenumber.  If the system is stable the apparent wavenumber is greater than the actual wavenumber and the interfacial waves oscillate more rapidly than they would in the absence of surface tension.  Conversely, when the system is unstable, the apparent wavenumber is smaller than the actual wavenumber and the growth rate is lower than in the absence of surface tension.  The growth rate is therefore inhibited by the surface tension.  We showed that there exists a critical wavenumber, $m^*$, above which the solutions to the dispersion relation are real and stable waves propagate about the interface, but below which the dispersion relation has complex conjugate pair solutions and so the most unstable mode must force growth of any perturbation.  In the special case $m=m^*$ the surface tension exactly balances the unstable density stratification and the perturbation is stationary.

We considered the effect of density diffusion at the interface between two miscible fluid layers.  Our motivation was to establish whether a diffuse interface changes the observed length scales and inhibits growth rates of the instability as it does in the classical gravity driven Rayleigh-Taylor and Kelvin-Helmholtz instabilities.  We considered two fluids with an initially sharp interface that was subject to a period of diffusion prior to the interface being perturbed.  We showed that if the density transition between the fluid layers can be accurately approximated by $\rho = \beta r^\alpha$ for two free constants $\alpha$ and $\beta$, then the Rayleigh equation still accepted power-law solutions and an analytical expression for the dispersion relation could be found.   Our finding is that the effect of the diffusion is always to inhibit the growth of any given mode. 

Full solutions to the Orr-Sommerfeld equation were also found in the case of fluids with differing, but uniform, densities and viscosities.  Here the solutions were of the form of power-laws and Bessel functions.  The increased number of eigensolutions is matched by an increase in the number of boundary conditions and interfacial continuity conditions.  In the limit of very high viscosity (low Reynolds number) the solutions behave like the Stokes flow, zero-inertia solutions, as they must.  In the limit of very low viscosity (high Reynolds number) the solutions behave like the inviscid solution, again, as they must.  The viscosity in the stable configuration always acts to slow down the speed of interfacial wave precession.   As the Reynolds number tends to zero, one of the two travelling wave solutions approaches the zero-inertia solution.  The other solution degenerates by its imaginary part tending to positive infinity.  In an unstable configuration we showed that while the decaying stable solution is sensitive to the viscosity contrast, the growth of the unstable mode is insensitive to the viscosity contrast.  The growth of the instability is due to the system Reynolds number and there is no strong dependence on whether it is the inner layer or the outer layer that is the most viscous.  In the limit of high viscosity (low Reynolds number) the solution behaves asymptotically like the zero-inertia solution and tends towards a stationary state.  The effects of diffusion of the interface or surface tension at the interface were also considered and shown to have qualitatively similar effects to the inviscid case; all cases may be considered analytically.  In particular there exists a mode above which surface tension is able to stabilize the growth of the perturbation.

Finally, the zero-inertia Stokes flow solutions were considered in comparison with established results for similar flows in Hele-Shaw cells and porous media.  The differences between the present configuration and the Hele-Shaw cell and porous media configurations lies in the way that viscosity acts on the flow.  The results differ by a factor that may be interpreted as a shape factor since the growth rates otherwise have identical physical dependencies on the Atwood and Weber numbers.

We have revisited well-established classical results in gravity-driven Rayleigh-Taylor instability but cast here in a rotating frame and with no gravity acting.  We have shown that it is possible to have Rayleigh-Taylor-like growth that is centrifugally driven.  The effects of surface tension, interface diffusion, and fluid viscosity are all seen to have qualitatively similar effects on the growth rate of the instability to the classical gravity-driven case.


\begin{acknowledgements}
MMS gratefully acknowledges useful discussions with Prof. J.~Billingham and Prof. J.~King.
\end{acknowledgements}


\appendix
\section{\label{sec:lrnsb}Low Reynolds number viscous solution behaviour}

We introduce the notation
\begin{equation}
\mathcal{J}_{i,j,k} = \mathcal{J}_{m+i}\left(\kappa^{1/2}\left[1 - (-1)^j\mathscr{A}\right]^{1/2}r_0^k\right), 
\quad
\mathcal{Y}_{i,j,k} = \mathcal{Y}_{m+i}\left(\kappa^{1/2}\left[1 - (-1)^j\mathscr{A}\right]^{1/2}r_0^k\right),  
\end{equation}
and the determinant
\begin{equation}
\mathcal{D}^{i_1, j_1, k_1}_{i_2, j_2, k_2} := \left|
\begin{array}{ll}
\mathcal{J}_{i_1, j_1, k_1} & \mathcal{J}_{i_2, j_2, k_2} \\
\mathcal{Y}_{i_1, j_1, k_1} & \mathcal{Y}_{i_2, j_2, k_2}
\end{array}\right|.
\end{equation}
For the negative root associated with $\omega_\infty^-$, we have eigenvalues $\omega \sim \lambda  + \textrm{i}\kappa\Rey^{-1}$ where $\kappa$ is a root of
\begin{multline}
\kappa^{1/2}\sqrt{1+\mathscr{A}}r_0 
\left(1+\mathscr{A}r_0^{2m}\right)
\left(\sqrt{1 + \mathscr{A}}\mathcal{J}_{1, 0, 1}\mathcal{D}^{0, 1, 1}_{1, 1, 0} + \sqrt{1 -  \mathscr{A}}\mathcal{J}_{0, 0, 1}\mathcal{D}^{1, 1, 0}_{1,  1, 1} \right)
\\ + 2m\mathscr{A}\Bigg\{\sqrt{1+\mathscr{A}}r_0^{2m + 1}\mathcal{J}_{1, 0, 1}\mathcal{D}^{0, 1, 0}_{0, 1, 1} \hspace*{200pt}
\\
-\sqrt{1-\mathscr{A}}\mathcal{J}_{0, 0, 1}
 \Bigg(\mathcal{D}^{0, 1, 1}_{1, 1, 0} + r_0^m\mathcal{D}^{1, 1, 0}_{0, 1, 0} + r_0^{m + 1}\mathcal{D}^{1, 1, 1}_{0, 1, 1} + r_0^{2m + 1}\mathcal{D}^{0, 1, 0}_{1, 1, 1} \Bigg) 
 \Bigg\}=0.
\end{multline}
The linear equation for the coefficient $\lambda$ in the expression for $\omega\sim\lambda + \kappa\textrm{i}\Rey^{-1}$ is given by
\begin{multline}\label{eq:imagAsy}
\Bigg\{2\kappa^2r_0^4\sqrt{1-\mathscr{A}}\mathcal{J}_{1, 0, 1}
	\Bigg[2\mathscr{A}^2
		\left(\sqrt{1+\mathscr{A}}\kappa^{3/2}\mathcal{D}^{1, 1, 0}_{1, 1, 1}+2m\kappa\mathcal{D}^{1, 1, 1}_{0, 1, 0}\right)r_0^{2m+1}
		\\
		+\mathscr{A}\bigg\{
			\left(2\sqrt{1+\mathscr{A}}m\kappa^{1/2}\mathcal{D}^{0, 1, 0}_{0, 1, 1}+\kappa(1+\mathscr{A})\mathcal{D}^{0, 1, 1}_{1, 1, 0}\right)(m+6)
			-(1+\mathscr{A})^{3/2}\kappa^{3/2}\mathcal{D}^{0, 1, 0}_{0, 1, 1}\bigg\}r_0^{2m}
		\\
		+\left(2\mathscr{A}r_0\mathcal{D}^{1, 1, 0}_{1, 1, 1}-(1+\mathscr{A})\mathcal{D}^{0, 1, 0}_{0, 1, 1}\right)\kappa^{3/2}\sqrt{1+\mathscr{A}}
		\\
		+\kappa\bigg\{2\mathscr{A}(1-\mathscr{A})mr_0^m\left(r_0\mathcal{D}^{1, 1, 1}_{0, 1, 1}+\mathcal{D}^{1, 1, 0}_{0, 1, 0}\right)-\left((2\mathscr{A}^2-\mathscr{A}+1)m
			-6(1+\mathscr{A})\right)\mathcal{D}^{0, 1, 1}_{1, 1, 0}\bigg\}\Bigg]
		\\
		-2\kappa^2\bigg[\mathscr{A}
			\left(-(m+6)\sqrt{1+\mathscr{A}}\kappa\mathcal{D}^{1, 1, 0}_{1, 1, 1}+\left((1+\mathscr{A})\kappa^{3/2}-2m(m+6)\sqrt{\kappa}\right)\mathcal{D}^{1, 1, 1}_{0, 1, 0}
			\right)r_0^{2m+4}
		\\
		+\kappa r_0^3\sqrt{1+\mathscr{A}}\left(2m\mathscr{A}(r_0\mathcal{D}^{1, 1, 0}_{1, 1, 1}-\mathcal{D}^{0, 1, 0}_{0, 1, 1})+r_0(m-6)\mathcal{D}^{1, 1, 0}_{1, 1, 1}\right)
		+r_0^4(1+\mathscr{A})\kappa^{3/2}\mathcal{D}^{1, 1, 1}_{0, 1, 0}
		\\
		+2m\mathscr{A}\kappa^{1/2}(m+6)r_0^3\left(\mathcal{D}^{1, 1, 0}_{0, 1, 0}r_0^m+\mathcal{D}^{1, 1, 1}_{0, 1, 1}r_0^{m+1}+\mathcal{D}^{0, 1, 1}_{1, 1, 0}\right)\bigg]	
		\mathcal{J}_{0, 0, 1}(1-\mathscr{A})\Bigg\}\lambda
\\
		-8\kappa^2r_0^3\mathscr{A}\Bigg\{\mathcal{J}_{0, 0, 1}(1-\mathscr{A})\bigg[\left(\sqrt{1+\mathscr{A}}\kappa\mathcal{D}^{1, 1, 0}_{1, 1, 1}+2\kappa^{1/2}m\mathcal{D}^{1, 1, 1}_{0, 1, 0}\right)r_0^{2m+1} - \kappa\sqrt{1+\mathscr{A}}r_0\mathcal{D}^{1, 1, 0}_{1, 1, 1}
	\\
	-2\kappa^{1/2}m\left(\mathcal{D}^{0, 1, 1}_{1, 1, 0} + r_0^m\mathcal{D}^{1, 1, 0}_{0, 1, 0}+r_0^{m+1}\mathcal{D}^{1, 1, 1}_{0, 1, 1}\right)\bigg]
	\\
	+\sqrt{1-\mathscr{A}}r_0\left[2r_0^{2m}\sqrt{1+\mathscr{A}}m\kappa^{1/2}\mathcal{D}^{0, 1, 0}_{0, 1, 1}-\kappa(1-r_0^{2m})(1+\mathscr{A})\mathcal{D}^{0, 1, 1}_{1, 1, 0}\right]\mathcal{J}_{1, 0, 1}\Bigg\}=0.
\end{multline}
An example solution is that for $m = 20$, $\mathscr{A} = 5/6$, $r_0 = 2^{-1/2}$, then $\kappa = 610.71$, $\lambda = -0.12$.



\end{document}